\documentclass[10pt,journal,compsoc]{IEEEtran} % Define IEEE TDSC template
\ifCLASSOPTIONcompsoc
  \usepackage[nocompress]{cite}
\else
  \usepackage{cite}
\fi

\usepackage{graphicx} 
\usepackage{minted} 

\usepackage{enumitem} 

\usepackage{xcolor}
\usepackage[caption=false]{subfig}
\usepackage{soul} % To highlight text
\definecolor{hlyellow}{RGB}{255,255,150}
\definecolor{hllightgreen}{RGB}{204,255,204}
\definecolor{hllightpink}{RGB}{255,200,210}

\newif\ifshownotes
\shownotestrue

\usepackage{tikz}
\newcommand{\numicon}[1]{%
  \tikz[baseline={([yshift=-0.7ex]current bounding box.center)}
] \node[draw,circle,
          inner sep=0pt,           % no extra padding
          line width=0.4pt,        % border thickness
          minimum size=0.8em,      % overall diameter
          font=\scriptsize         % text size inside
         ](char){#1};%
}

\usepackage{tabularx}
\usepackage{booktabs}
\usepackage{threeparttable}
\usepackage{multirow}

\usepackage{adjustbox}

\usepackage{siunitx}
\usepackage[ruled,vlined]{algorithm2e}
\usepackage{amsmath}

\usepackage{amssymb} 
\usepackage{makecell}
\usepackage{array}
\usepackage{xspace}
\usepackage[table]{xcolor}
\definecolor{cfirow}{gray}{0.92}
\usepackage{wasysym} 

\usepackage{pifont}

\newcommand{\SolidSquare}{$\blacksquare$}
\newcommand{\WhiteSquare}{$\square$} 
 
\newcommand{\FullMark}{\CIRCLE}
\newcommand{\Present}{\SolidSquare}
\newcommand{\NotPresent}{\WhiteSquare}
\newcommand{\LimitedMark}{\ding{115}}
\newcommand{\KscfaMark}{\ding{72}}
\newcommand{\GedenMark}{\ding{67}}

\newcommand{\PartialMark}{\LEFTcircle}
\newcommand{\CheckMark}{$\checkmark$}
\newcommand{\schemename}{KS-CFA\xspace}
\newcommand{\recordInputVar}{ISCFD\xspace}
\newcommand{\CFInputVar}{ICFD\xspace}

\usepackage{xspace}
\usepackage{pifont}
\newcommand{\xmark}{\ding{55}}
\usepackage{hyperref}

\newcommand{\verifier}{$\mathcal{V}$\xspace} %Verifier
\newcommand{\prover}{$\mathcal{P}$\xspace}   %Prover

\begin{document}

\title{KS-CFA: Control-Flow Attestation via Symbolic Replay Against Control-Flow Bending Attacks}
\author{Zhanyu Sha, Konstantinos Markantonakis, Carlton Shepherd, Amir Rafi
\thanks{Z.\ Sha, K.\ Markantonakis and A.\ Rafi are with the Information Security Group,
Royal Holloway, University of London, Egham, Surrey, United Kingdom. Corresponding author: Z.\ Sha (e-mail: zhanyu.sha.2022@rhul.ac.uk)}
\thanks{C.\ Shepherd is with the Department of Computer Science, Durham University, Durham, UK, and was supported by the UK Engineering and Physical Sciences Research Council through grant EP/Y030168/1.}
}

\maketitle

\begin{abstract}
Control-flow attestation (CFA) enables a remote entity to verify program execution on a target device by monitoring control-flow behaviour at runtime. However, control-flow bending (CFB) attacks remain difficult to detect, where an adversary steers execution along legal edges of the program's control-flow graph by corrupting branch flags, loop counters, and other runtime data. 
Existing solutions impose significant drawbacks: they require enumerating vast measurement spaces, cover only a reduced subset of attacks, or rely on low-level hardware modifications. 
In this work, we present \schemename, a new CFA scheme that detects CFB attacks across four transfer types (indirect calls, conditional and indirect jumps, and returns) without those costs. To this end, we combine symbolic execution and selective identification of input-sourced control-flow dependent variables: a strict subset of control-flow-relevant state whose values are directly read from external input. The proving device records, inside a trusted execution environment (TEE), a control-flow trace and the external inputs that determine relevant run-time variables. The verifier then replays the reported path through single-path symbolic execution, predicting each transfer and localising divergences that signal an attack. We implement and evaluate \schemename using the RISC-V Keystone TEE and Embench-IoT on QEMU and a Rocket-based FPGA platform (NiteFury~II). Prover-side overhead relative to unattested execution ranges from 6.8--20.5$\times$ on QEMU and 6.7--32.2$\times$ on the FPGA, and verification requires no path or value enumeration.
\end{abstract}

\section{Introduction}
\IEEEPARstart{C}{ontrol-flow} attacks remain a persistent and powerful threat to system security. Traditional control-flow attacks (e.g., code injection, code reuse) tamper with control data (e.g., return addresses, function pointers) to redirect execution through memory-corruption vulnerabilities. Such attacks have been mitigated by control-flow integrity (CFI) techniques, which enforce that a program's runtime execution follows its control-flow graph (CFG). CFI operates locally on the device and cannot provide remote, cryptographically verifiable evidence of execution correctness. This gap has motivated the emergence of control-flow attestation (CFA)~\cite{C-FLAT,DO-RA,LiteHAX,ACFA,Atrium,CHASE,ben2022nanovised,geden2019hardware,GuaranTEE,LAPE,LightFAt,LO-FAT,OAT,RADIS,RAGE,RECFA,BLAST,Tiny-CFA,SCARR,CFASoKPaper,BDFCFA,CFRV,ISC-FLAT,Log-based_CFA,MGC-FA,ARI}, which enables a remote verifier (\verifier{}) to validate a program's execution path against attestation reports sent from the prover device (\prover{}).

However, CFA does not detect all control-flow attacks. A particularly subtle class are \emph{control-flow bending} (CFB) attacks~\cite{CFB,Non-control-data-attacks}, which are fundamentally harder to detect than traditional control-flow threats, such as return-oriented programming (ROP)~\cite{ROP_Original}. In a CFB attack, the adversary manipulates control- and non-control data---branch flags, loop counters, indirect-call targets, or return addresses stored in writable memory---to steer execution along edges that are legal in the CFG, but produce malicious behaviour collectively (e.g., bypassing security checks or escalating privileges). Because individual control-flow transfers respect the CFG, detecting such attacks requires reasoning about whether the runtime values that drove control-flow decisions were themselves legitimate.

Few CFA proposals reason about these values at all; those that do provide only partial CFB coverage or do so at substantial cost. DO-RA~\cite{DO-RA} extends the conventional approach of hash-based attestation---where cumulative hashes are computed at successive nodes of a target program's CFG (see C-FLAT~\cite{C-FLAT})---by folding the values of runtime variables into the digest. This obligates \verifier{} to pre-enumerate the set of all legitimate hashes, which is intractable when control flow depends on continuous or high-cardinality inputs. 
BDFCFA~\cite{BDFCFA} instead requires \verifier{} to supply the program's inputs, 
which presumes \verifier{} knows the complete space of program inputs and fails when inputs originate from unpredictable sources (e.g., sensor feeds, system state). Geden et al.~\cite{geden2019hardware} monitors the legality of individual control-flow transfers in hardware, requiring bespoke CPU pipeline extensions that report only a violation flag rather than detailed traces. 
Other schemes cover CFB only partially: 
OAT~\cite{OAT} and ARI~\cite{ARI} monitor control-flow-dependent variables at  runtime, but only a subset of them, yielding partial coverage within each transfer type, 
while LiteHAX~\cite{LiteHAX} cannot distinguish a bending attack from a legitimate access within the same memory allocation.

In this paper, we present \schemename, a CFA scheme that closes the gap between CFB coverage and the usual costs of attaining it. 
Our design rests on two ideas. The first is  
\emph{selective input recording}. 
In an offline phase, \verifier{} analyses the target program at the LLVM~intermediate representation (IR) level and classifies every value as either \emph{verifier-known} (determinable from code, constants, and compile-time facts) or \emph{verifier-unknown} (dependent on runtime inputs or environment state). Among verifier-unknown values, those that are directly read from external input and that influence control-flow decisions (e.g., branch conditions, loop bounds, call/jump targets), either directly or through data dependencies, are called \emph{input-sourced control-flow dependent} (\recordInputVar) variables. At runtime, when an \recordInputVar variable is assigned by an external input, \prover{} records the data read at that assignment and includes it in the attestation report. The second idea is \emph{single-path symbolic replay}. Rather than pre-computing hashes for every CFG path combination, \verifier{} validates \prover's attestation report---a sequence of control-flow transfers---by replaying the reported path. During replay, \verifier propagates known values symbolically and substitutes the recorded inputs at the corresponding \recordInputVar assignments. \verifier checks that each branch decision is semantically consistent with the reported transfer. These two mechanisms together let \verifier combine program semantics with inputs in order to identify semantic infeasibility where the reported successor diverges from \verifier's prediction.

We thus contribute a CFA proposal that achieves wide CFB coverage without \verifier-side path or value enumeration. KS-CFA covers four exploitable transfer types (indirect calls, conditional jumps, indirect jumps, and returns) and requires a standard trusted execution environment (TEE) without additional hardware. 
We implement \schemename{} on a RISC-V platform using the Keystone TEE and evaluate it under both QEMU and a NiteFury~II FPGA board.  The remainder of this paper is organised as follows. \S\ref{sec:background} covers background and related work. \S\ref{sec:threat-model} defines the threat model and assumptions. \S\ref{System overview} and \S\ref{KS-CFA Design} present the system overview and detailed design, respectively. \S\ref{sec:implementation} describes the implementation, \S\ref{sec:evaluation} evaluates overhead and verification effectiveness, 
and \S\ref{sec:security-analysis} presents a security analysis. \S\ref{sec:limitations} discusses limitations and future work before concluding in \S\ref{sec:conclusion}.

\section{Background and Related Work} \label{sec:background}

We introduce control-flow integrity, attestation, and bending attacks, and identify the limitations of existing work.

%% ============================================================

\subsection{Control-Flow Integrity and Attestation} \label{sec:cfi-cfa}

CFI~\cite{CFI} enforces that every runtime control-flow transfer respects a predefined policy, typically expressed as a CFG. A CFG is a directed graph $G=(V,E)$ whose vertices $v \in V$ represent basic blocks (BBLs) and whose edges $(v_i, v_j) \in E$ represent permitted transfers. A program satisfies CFI if every runtime transfer follows an edge in~$E$. CFGs are derived by \emph{static analysis} of source code, binaries, or an intermediate representation such as LLVM~IR, or by \emph{dynamic analysis} of observed test executions. We denote a control-flow path as the sequence of BBLs visited in a single run, $T=(v_{1},\dots,v_{n})$, equivalently the edge list $\{(v_{1},v_{2}),\dots,(v_{n-1},v_{n})\}$.

CFA~\cite{CFASoKPaper} addresses CFI's local-only limitation by producing an authenticated record of control-flow behaviour. A common design pattern involves \verifier{} issuing a challenge to \prover{}, which then captures control-flow events during execution. \prover{} hashes control-flow events cumulatively, returning a report that is authenticated by a trust anchor. \verifier{} validates and verifies the report using a pre-computed database of legitimate hashes derived from the program's CFG, flagging any deviations. CFA proposals have varied, firstly, around \emph{what is measured}: cumulative hashes at CFG nodes (C-FLAT~\cite{C-FLAT}, Atrium~\cite{Atrium}); explicit edge or path sequences (ReCFA~\cite{RECFA}, OAT~\cite{OAT}, BLAST~\cite{BLAST}); and measurements augmented with data values (DO-RA~\cite{DO-RA}, ARI~\cite{ARI}) or memory-access metadata (LiteHAX~\cite{LiteHAX}). Secondly, proposals vary according to \emph{how} \verifier{} validates the report: hash matching against a pre-computed database (C-FLAT~\cite{C-FLAT}, DO-RA~\cite{DO-RA}); CFG-membership over an explicit trace (ReCFA~\cite{RECFA}, OAT~\cite{OAT}); value-based checks on instrumented variables (OAT~\cite{OAT}, ARI~\cite{ARI}); or learned models over trace embeddings (RAGE~\cite{RAGE}). Proposals also vary by trust anchor, such as TEEs (ARM TrustZone~\cite{C-FLAT,OAT,DO-RA} and Intel SGX~\cite{BDFCFA, GuaranTEE}) or dedicated co-processors and on-chip extensions~\cite{LO-FAT,LiteHAX,CHASE,geden2019hardware}. The reader is referred to Sha et al.~\cite{CFASoKPaper} for a survey of CFA schemes.

\subsection{Control-Flow Bending Attacks} \label{sec:cfb}

\begin{figure}[t]
    \centering
    \includegraphics[width=\columnwidth]{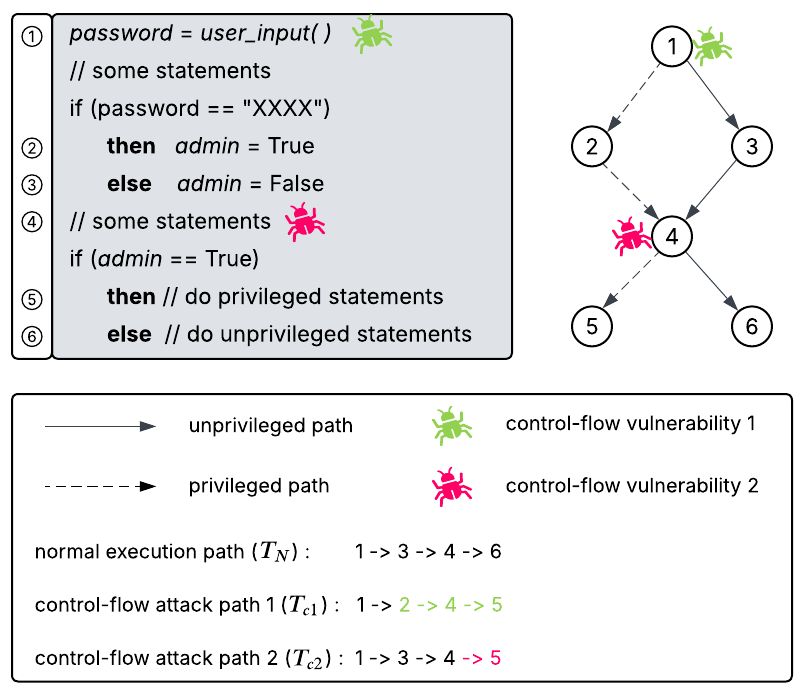}
    \caption{A program's CFG and two CFB attacks. $T_N$ is the benign (unprivileged) path; $T_{c1}=(B_1,B_2,B_4,B_5)$ and  $T_{c2}=(B_1,B_3,B_4,B_5)$ are compromised paths whose edges are CFG-legal. (Derived from~\cite{debes2023zekra}).}
    \label{fig:simple CFG}
\end{figure}

In a CFB attack, the adversary corrupts non-control data (branch flags, loop counters) or control data (function pointers, return addresses) to produce a legal but malicious path. Such attacks have been demonstrated against real-world production software (e.g., CVE-2013-2028~\cite{CFB}). Fig.~\ref{fig:simple CFG} gives an example program that reads a password, sets a Boolean flag, and branches on it. CFG $G_1$ has vertices $V=\{B_1,\dots,B_6\}$ and edges $E=\{(B_1,B_2),(B_1,B_3),\dots,(B_4,B_6)\}$. Under benign execution with an incorrect password, the program follows $T_N=(B_1,B_3,B_4,B_6)$ and performs unprivileged operations. Two CFB attacks can achieve privilege escalation:
\begin{enumerate}
    \item \emph{Password overwrite.} The attacker is able to exploit a vulnerability in $B_1$ to read the hard-coded constant \texttt{"XXXX"} and overwrite the \texttt{password} variable, forcing path $T_{c1}=(B_1,B_2,B_4,B_5)$.
    \item \emph{Flag tampering.} The attacker is able to corrupt the \texttt{admin} variable after $B_3$ executes, forcing path $T_{c2}=(B_1,B_3,B_4,B_5)$.
\end{enumerate}
Every edge in both $T_{c1}$ and $T_{c2}$ appears in $G_1$. A dynamic CFG constructed from comprehensive test inputs may include $T_{c1}$---since it coincides with the legitimate path when the correct password is supplied---but could exclude $T_{c2}$ if no test input causes $B_3$ to be followed by $B_5$. 
Generally, when a CFB attack follows a path that is reachable under some benign input, no CFG-based check---static or dynamic---can detect it.  Detecting such attacks requires reasoning about runtime data beyond the legality of edges.

%% ============================================================
\subsection{CFB Detection Approaches} \label{sec:existing-approaches}

Traditional CFA schemes such as C-FLAT~\cite{C-FLAT}, LO-FAT~\cite{LO-FAT}, and CHASE~\cite{CHASE} use CFGs to detect control-flow attacks but do not reason about control-flow data. CFB attacks are not detected with static CFGs because the paths are CFG-legal (although call-return matching can still detect CFB attacks on returns~\cite{C-FLAT}). A dynamic CFG detects CFB attacks on returns because a corrupted return is not an observed path. Nevertheless, only attacks on unobserved paths are caught, missing those that coincide with legitimate executions.

To address this, OAT~\cite{OAT} uses value-based checking on control-flow dependent variables, but restricts protection to developer-annotated code regions. ARI~\cite{ARI} uses data-flow protection for critical variables in targeted regions and reports their values to \verifier, achieving broader but still incomplete coverage. LiteHAX~\cite{LiteHAX} reports runtime memory-access metadata (addresses of loads, stores, and the memory locations they touch) to \verifier, enabling detection of CFB attacks arising from memory corruptions across distinct data objects. However, when the corruption occurs in the same allocation as the overflowed buffer (e.g., fields of the same \texttt{struct}), address-level checks cannot distinguish between corruption and legitimate accesses. RAGE~\cite{RAGE} applies graph neural networks to control-flow traces, reporting F1-scores of 98.03\% for ROP and 91.01\% for data-oriented programming (DOP) attacks on Embench-IoT. Detection is probabilistic due to the learning paradigm; a model trained on benign traces cannot reliably distinguish CFB attacks whose paths resemble legitimate executions (e.g., $T_{c1}$ in Fig~\ref{fig:simple CFG}).

In the CFI literature, context-sensitive schemes refine the set of legal targets for each indirect control-flow transfer. For such a given transfer, the \emph{equivalence class} (EC) is the set of target BBLs that the mechanism cannot distinguish using only the static CFG; a smaller EC implies stronger enforcement. PITTYPAT~\cite{PITTYPAT}, $\mu$CFI~\cite{uCFI}, CFI-LB~\cite{CFI-LB}, and OS-CFI~\cite{OS-CFI} reduce EC sizes for indirect calls and returns (typically protecting returns via shadow stacks) but do not achieve singleton ECs for all indirect control-flow transfers, providing only partial protection. BCI-CFI~\cite{BCI-CFI} addresses indirect jumps but does not cover conditional branches.

vCFI~\cite{vCFI} follows a maximalist approach by protecting \emph{all} control-flow dependent variables through shadow memory-based integrity checking; however, it does not support remote attestation. DO-RA~\cite{DO-RA} extends attestation measurements to encompass the runtime values of control-flow dependent variables and data-flow dependencies. Offline, \verifier locates branches and loops, identifies constraining variables, and rewrites the binary with stubs that record transfers and variable values. At runtime, \prover hashes the recorded sequence, and \verifier checks the result against a pre-constructed database of legitimate hashes. Because the hash embeds exact runtime values, each distinct value produces a different digest. When control-flow depends on continuous input(s), \verifier must pre-enumerate a large set of legitimate digests, which is intractable for programs reliant on high-cardinality or continuous-valued inputs.

BDFCFA~\cite{BDFCFA} incorporates expected input values by recording hashes of legal paths with the input ranges that produce them. \prover reports a path hash; \verifier validates it against inputs that \verifier itself supplied. This model requires \verifier to be knowledgeable of all program inputs, which lacks scalability when inputs originate from unpredictable sources (e.g., configuration files, system variables, sensor feeds).  Geden et al.~\cite{geden2019hardware} uses a dedicated hardware module in the CPU pipeline that checks control-flow dependent variables. It performs local attestation, reporting violation flags with limited diagnostic information (e.g., the last executed instruction), limiting \verifier-side attack traceability. SABRE \cite{SABRE} performs CFG-based path verification followed by automatic root-cause analysis and binary patching for buffer overflow and use-after-free vulnerabilities.

\begin{table}[t]
  \centering
  \caption{Comparative summary of related CFI (\colorbox{cfirow}{\strut shaded rows}) and CFA (unshaded) schemes.}
  \label{tab:general comparison}
  \begin{adjustbox}{max width=\columnwidth}
  \begin{threeparttable}
        \small
        \begin{tabular}{%
            l   % Scheme
            c   % Input type
            c   % A1
            c   % A2
            c   % A3
            c   % A4
            c   % Traceable
            c   % Dependencies
        }
          \toprule
          \textbf{Scheme}
            & \makecell[c]{\textbf{Input}\\\textbf{type}}
            & \textbf{A1}
            & \textbf{A2}
            & \textbf{A3}
            & \textbf{A4}
            & \makecell[c]{\textbf{Trace-}\\\textbf{able}}
            & \textbf{Deps.}
            \\
          \midrule

          \rowcolor{cfirow}
          PathArmor~\cite{PathArmor}
            & Binary
            & \PartialMark & \PartialMark & \xmark & \xmark
            & ---
            & Intel LBR \\

          \rowcolor{cfirow}
          PITTYPAT~\cite{PITTYPAT}
            & LLVM
            & \PartialMark & \FullMark & \xmark & \xmark
            & ---
            & Intel PT \\

          \rowcolor{cfirow}
          $\mu$CFI~\cite{uCFI}
            & LLVM
            & \PartialMark & \FullMark & \xmark & \xmark
            & ---
            & Intel PT \\

          \rowcolor{cfirow}
          CFI-LB~\cite{CFI-LB}
            & LLVM
            & \PartialMark & \FullMark & \xmark & \xmark
            & ---
            & Intel TSX \\

          \rowcolor{cfirow}
          OS-CFI~\cite{OS-CFI}
            & LLVM
            & \PartialMark & \FullMark & \xmark & \xmark
            & ---
            & Intel TSX \\

          \rowcolor{cfirow}
          BCI-CFI~\cite{BCI-CFI}
            & LLVM
            & \PartialMark & \FullMark & \PartialMark & \xmark
            & ---
            & SW \\

          \rowcolor{cfirow}
          vCFI~\cite{vCFI}
            & LLVM
            & \FullMark & \FullMark & \FullMark & \FullMark
            & ---
            & SW \\

          \midrule

          C-FLAT~\cite{C-FLAT}
            & Binary
            & \xmark & \FullMark & \xmark & \xmark
            & \xmark
            & TrustZone \\

          LiteHAX~\cite{LiteHAX}
            & Source
            & \PartialMark & \FullMark & \PartialMark & \PartialMark
            & \LimitedMark
            & Custom CPU HW \\

          OAT~\cite{OAT}
            & LLVM
            & \PartialMark & \PartialMark & \PartialMark & \PartialMark
            & \LimitedMark
            & TrustZone \\

          ARI~\cite{ARI}
            & LLVM
            & \PartialMark & \PartialMark & \PartialMark & \PartialMark
            & \LimitedMark
            & TrustZone \\

          Geden et al.~\cite{geden2019hardware}
            & Binary
            & \FullMark & \FullMark & \FullMark & \FullMark
            & \LimitedMark
            & Custom CPU HW\\

          DO-RA~\cite{DO-RA}
            & Binary
            & \FullMark & \FullMark & \FullMark & \FullMark
            & \xmark
            & TrustZone \\

          BDFCFA~\cite{BDFCFA}
            & Binary
            & \FullMark & \FullMark & \FullMark & \FullMark
            & \xmark
            & Intel SGX \\

          \midrule

          \textbf{KS-CFA}
            & LLVM
            & \FullMark & \FullMark & \FullMark & \FullMark
            & \FullMark
            & TEE \\
          \bottomrule
        \end{tabular}

        \vspace{0.5ex}

        \begin{tablenotes}
          \item A1: CFB-ICALL; A2: CFB-RET; A3: CFB-IJMP; A4: CFB-CJMP.
          \item \PartialMark~Partial coverage; \FullMark~Full coverage; \xmark~No coverage.
          \item \LimitedMark:  (OAT~\cite{OAT}, ARI~\cite{ARI}: marked regions only;
                Geden et al.~\cite{geden2019hardware}: violation flag and last executed instruction; LiteHAX~\cite{LiteHAX}: control-flow transfers reported as runtime addresses,
                but load/stores only as running hashes).
          \item ---~Not applicable: CFI enforces locally and produces no report for \verifier{}.
          \item Deps.\ = platform dependencies; HW = hardware; SW = software-based.
        \end{tablenotes}
  \end{threeparttable}
  \end{adjustbox}
\end{table}
\begin{table}
  \centering
  \caption{Comparison of CFA schemes with comprehensive CFB detection.}
  \label{tab:dedicated-cfa-comparison}
  \resizebox{0.95\linewidth}{!}{%
  \begin{threeparttable}
  \small
  \begin{tabular}{@{} l c c c c c c c @{}}
  \toprule
  \textbf{Scheme}
    & \makecell{\textbf{RT}\\\textbf{Addr}}
    & \makecell{\textbf{Var}\\\textbf{Vals}}
    & \makecell{\textbf{Ext}\\\textbf{Inp}}
    & \makecell{\textbf{Viol}\\\textbf{Flags}}
    & \makecell{\textbf{Vars}\\\textbf{Prot}}
    & \makecell{\textbf{Pre-}\\\textbf{enum}}
    & \makecell{\textbf{Ded}\\\textbf{HW}} \\
  \midrule
  DO-RA~\cite{DO-RA}
    & \Present & \Present & \NotPresent & \NotPresent & \NotPresent & \Present & \NotPresent \\
  BDFCFA~\cite{BDFCFA}
    & \Present & \NotPresent & \Present & \NotPresent & \NotPresent & \Present & \NotPresent \\
  Geden et al.~\cite{geden2019hardware}
    & \NotPresent & \NotPresent & \NotPresent & \Present & \GedenMark & \NotPresent & \Present \\
  \textbf{KS-CFA}
    & \Present & \NotPresent & \KscfaMark & \NotPresent & \NotPresent & \NotPresent & \NotPresent \\
  \bottomrule
  \end{tabular}
  \begin{tablenotes}
    \item \Present:~Property present; \NotPresent:~Absent.
    \GedenMark:~All control-flow dependent variables (\textbf{V.1});
    \KscfaMark:~Limited to control-flow-influencing inputs that \verifier{} does not supply (a strict subset of all program inputs). RT Addr = runtime addresses in measurement; Var Vals = variable values; Ext Inp = external inputs;
    Viol Flags = attacks reported as violation flags without execution trace;
    Vars Prot = variables protected at runtime; Pre-enum = \verifier{} must pre-enumerate legitimate measurements; Ded HW = dedicated hardware required.
  \end{tablenotes}
  \end{threeparttable}
  }
\end{table}

Tables~\ref{tab:general comparison} and~\ref{tab:dedicated-cfa-comparison} summarise our comparison. Few schemes detect CFB attacks across four exploitable transfer types: indirect calls (CFB-ICALL), conditional direct jumps (CFB-CJMP), indirect jumps (CFB-IJMP), and returns (CFB-RET) (defined in \S\ref{sec:threat-model}). Existing CFB-aware schemes reveal a tension between attack detection coverage and deployability. Hash-based schemes require \verifier{} to pre-enumerate the set of legitimate measurement values during the offline phase, which lacks scalability. 
Dedicated hardware approaches~\cite{geden2019hardware,CHASE} obviate software overhead but require custom low-level modifications.
\schemename{} addresses this gap by coupling selective recording of salient inputs that influence control-flow decisions (\S\ref{sec:iscfd}), a subset of all program inputs, with \emph{single-path symbolic replay} (\S\ref{sec:abstract-exec}). This combination gives wide CFB coverage across the four transfer types and provides detailed attack traceability. It avoids \verifier-side enumeration of legitimate paths, 
the requirement that \verifier be the sole source of program inputs, 
and dependencies on dedicated hardware beyond a standard TEE.

\section{Threat Model \& Assumptions} \label{sec:threat-model}

In this section, we discuss the threat model and assumptions of our proposed scheme.

\subsection{Attacker Model}
\label{sec:attacker-model}
We assume a software adversary who can exploit memory-corruption vulnerabilities (e.g., buffer overflows, use-after-free) to perform arbitrary reads and writes in the application's address space, including stack, heap, and global data. The attacker possesses full knowledge of the program's source code and CFG, and can approximate the dynamic CFG by observing sufficiently many test executions, enabling CFB attacks.  Further, we assume that the target system enforces write-xor-execute (W$\oplus$X) memory protection, where writable memory is non-executable and executable memory is non-writable. Consequently, the attacker cannot introduce new instructions when this is enabled; any register manipulation must use instructions already present in the program text. This is a standard assumption in the CFI and CFA 
literature~\cite{MCFI,C-FLAT,CHASE,DO-RA,LiteHAX,RECFA,CFASoKPaper}.

\verifier is assumed to be trusted and possesses the program's intended source code. \verifier receives an authenticated copy of the target program in the offline phase, and the binary is securely deployed on \prover. We rely on a GlobalPlatform-style TEE with secure and non-secure worlds separated with hardware-enforced isolation such that an attacker cannot read or modify the state of secure world-resident software from the untrusted world~\cite{shepherd2024trusted}.  The following are out of scope: attacks targeting the TEE itself (e.g., microarchitectural side channels, fault injection), physical attacks, denial-of-service, and data-only attacks (e.g., DOP attacks~\cite{DOP}) that do not change the control-flow path.

\subsection{Targeted Attack Classes} \label{sec:attack-classes}

Control-flow attacks exploit transfers whose concrete target can vary at runtime. Following prior systematisations~\cite{CFASoKPaper}, KS-CFA targets the four transfer types that can be weaponised for CFB attacks: 

\begin{itemize}
    \item \textbf{CFB-ICALL.} The adversary corrupts a function pointer or virtual-table entry so that an indirect call (\texttt{icall}) reaches an unintended yet CFG-legal callee.
    \item \textbf{CFB-CJMP.} The adversary manipulates a branch predicate or loop counter to flip the outcome of a conditional direct jump (\texttt{cjmp}) or alter iteration counts, steering execution down a legal but unintended path.
    \item \textbf{CFB-IJMP.} The adversary tampers with an index into a jump table (or another pointer used by an indirect branch/jump (\texttt{ijmp}), redirecting execution to a sibling BBL without violating the CFG.
    \item \textbf{CFB-RET.} The adversary overwrites a saved return address so a callee returns (\texttt{ret}) to a different but CFG-legal call site, breaking call-return correspondence.
\end{itemize}

Unconditional direct jumps (\texttt{ujmp}) and direct calls (\texttt{dcall}) have exactly one statically encoded destination; no attacker-controlled data determines the edge, so neither CFB nor traditional control-flow attacks can leverage them. KS-CFA also detects traditional control-flow attacks, such as ROP~\cite{ROP_Original} and call- (COP)~\cite{COP} and jump-oriented programming (JOP)~\cite{bletsch2011jump}, which produce transfers absent from the CFG, as these are strictly easier to identify than CFB attacks.

% =============================================================
\section{System Overview} \label{System overview}
% =============================================================

\begin{figure*}[t]
    \centering
    \includegraphics[width=0.76\linewidth]{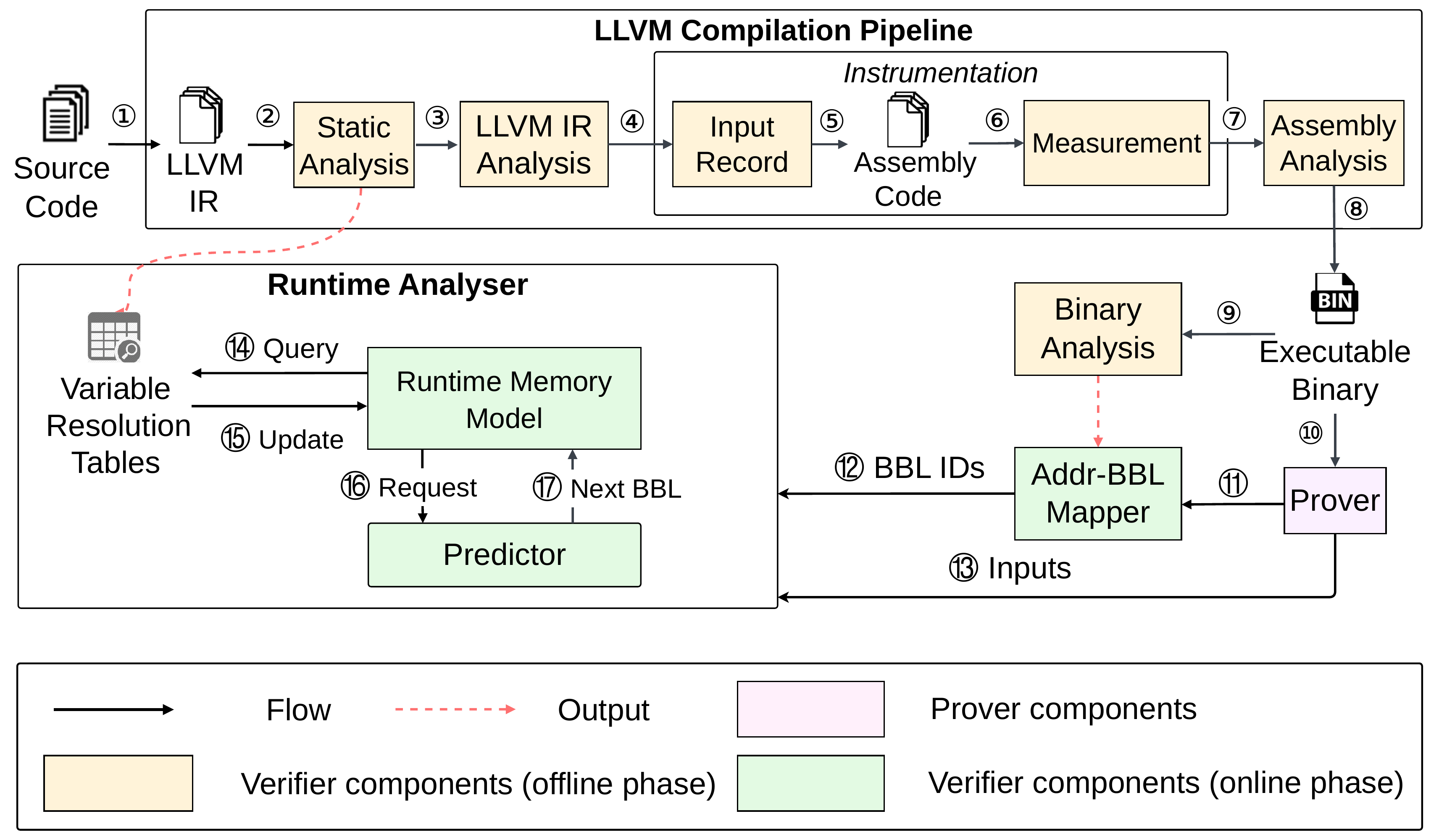}
    \caption{Overview of KS-CFA. (Circled steps are described in the text).}
    \label{KS-CFA workflow}
\end{figure*}

We now describe our solution. Fig.~\ref{KS-CFA workflow} illustrates the high-level architecture of KS-CFA, which follows two phases: an \emph{offline phase} (compilation and instrumentation) and an \emph{online phase} (attestation and verification).

In an offline phase, \verifier compiles the target program to a RISC-V binary through the LLVM toolchain. During compilation it performs static analysis and instrumentation on LLVM~IR and assembly files~\numicon{1}--\numicon{7} before producing the executable binary to be deployed on \prover{} in~\numicon{8}. Static analysis determines, for every BBL, which variable values can be resolved from the source code alone and which depend on runtime inputs. The results are stored in \emph{Variable Resolution Tables}~(VRTs) that \verifier later consults during verification. After compilation, binary analysis~\numicon{9} constructs an \emph{Addr-BBL mapper} that translates runtime addresses and branch outcomes back to LLVM~IR BBL identifiers. The final executable is deployed on \prover~\numicon{10}.

The online phase is at the point of execution on \prover. After the target program runs and terminates, \prover transmits the captured control-flow path and input log to \verifier. 
The Addr-BBL mapper receives the path~\numicon{11} and translates it into a sequence of LLVM~IR BBL identifiers~\numicon{12}. \verifier performs abstract execution~\numicon{13} using the BBL sequence and the input log: a step-by-step symbolic replay of the reported path, querying the VRTs~\numicon{14} to update a \emph{runtime memory model}~\numicon{15} and invoking a \emph{predictor} to determine the legitimate successor at control-flow transfers~\numicon{16}--\numicon{17}. A mismatch between the predicted and reported successor signals an attack.

\begin{figure}[t]
    \centering
    \includegraphics[width=\columnwidth]{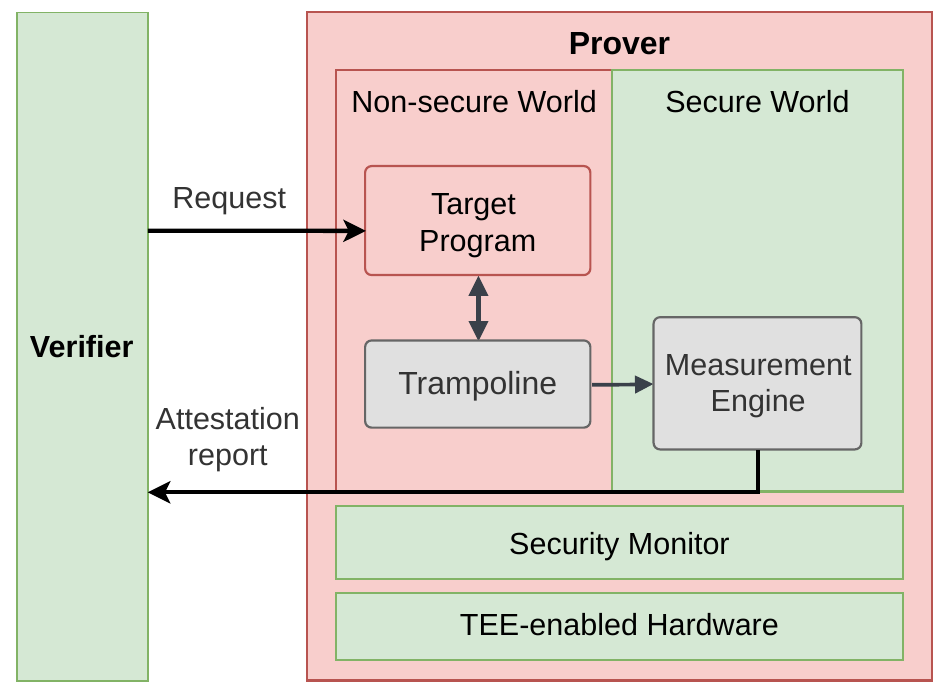}
    \caption{High-level architecture. Grey indicates components introduced by KS-CFA; green and red show trusted and untrusted areas respectively.}
    \label{Prover figure}
\end{figure}

Fig.~\ref{Prover figure} illustrates \prover's system architecture. The target program and a \emph{trampoline} execute in the non-secure world; a \emph{measurement engine} executes inside a secure enclave. When \verifier initiates attestation, \prover begins executing the target program. At every designated control-flow transfer, the trampoline intercepts execution and forwards the transfer information to the measurement engine in the TEE, which appends it to a control-flow log. 
When an assignment brings an external input that \verifier has not supplied into a variable whose value directly or indirectly influences a control-flow decision (\S\ref{sec:iscfd}), the trampoline forwards the data read at that assignment to the measurement engine, which appends it to an input log.
After termination, the signed logs are sealed into an attestation report and transmitted to \verifier.  The transport scheme is orthogonal to \schemename{}: any remote-attestation protocol ensuring
authenticity, integrity, and freshness of the report contents suffices (e.g.,~\cite{weinhold2025separate,KeystoneTEE,johnson2021taxonomy}).

% =============================================================
\section{KS-CFA Design} \label{KS-CFA Design}
% =============================================================

The central idea is to enable CFA without pre-enumerating legal paths. Instead of building a CFG or hashing all valid traces offline, \verifier replays the single path reported by \prover, using pre-computed program semantics to predict what a legitimate execution should do at each control-flow decision.  
When \verifier can resolve the relevant variables using program semantics, it predicts the successor directly and catches any deviation. When it cannot---because the decision depends on program inputs---\verifier has those inputs (either provided by \verifier or recorded by \prover in the input log) and uses them to resolve the relevant variables.
We present the design in three parts: the \recordInputVar variable concept that determines \verifier's knowledge boundary (\S\ref{sec:iscfd}), the abstract-execution procedure that is the core verification mechanism (\S\ref{sec:abstract-exec}), and the measurement infrastructure that captures and translates \prover's control-flow path (\S\ref{sec:measurement}).

\subsection{\recordInputVar Variables and Input Recording} \label{sec:iscfd}
KS-CFA requires program inputs because not all predictions can succeed through program semantics alone. However, \verifier does not need every program input, only those that influence control-flow decisions. The following conditions identify the target variables:

\begin{enumerate}[leftmargin=*, labelwidth=*, align=left]
    \item[\textbf{[V.1]}:] Its value can directly or indirectly (e.g., through data dependency) affect a control-flow transfer (e.g.\ a branch predicate, a loop counter, or a function pointer),
    \item[\textbf{[V.2]}:] Its value is \emph{data-dependent} on external input---e.g., user input, environment variables, files, sockets, system calls, or return values from third-party libraries---rather than being determined by program constants alone.
    \item[\textbf{[V.2a]}:] A strict subset of \textbf{[V.2]}. The variable has at least one assignment that takes its value directly from an external-input read.
\end{enumerate}

\begin{figure}[t]
\centering
  
  \begin{minted}[
    breaklines,
    autogobble,
    fontsize=\small,
    frame=lines,
    framesep=4pt,
  ]{c}
int a = user_input(); // external input
int b = a + 3;    // data-flow dependent on a 
bool c;
int d = user_input(); // independent input 

if (b > 0) { c = true; } 
else       { c = false; }

if (c) { /* ... */ }
else   { /* ... */ }
  \end{minted}
  \smallskip
  \centering\small

  % ── Invariant classification ──────────────────────────────────────────────
  \begin{threeparttable}
    \begin{tabular}{@{} l *{5}{c} @{}}
      \toprule
      \textbf{Var.}
        & \textbf{[V.1]}
        & \textbf{[V.2]}
        & \textbf{[V.2a]}
        & \textbf{\CFInputVar}
        & \textbf{\recordInputVar} \\
      \midrule
      $a$ & \CheckMark & \CheckMark & \CheckMark & \CheckMark & \CheckMark \\
      $b$ & \CheckMark & \CheckMark & \xmark     & \CheckMark & \xmark     \\
      $c$ & \CheckMark & \xmark     & \xmark     & \xmark     & \xmark     \\
      $d$ & \xmark     & \CheckMark & \CheckMark & \xmark     & \xmark     \\
      \bottomrule
    \end{tabular}
    \begin{tablenotes}\small\upshape
      \item \CheckMark~=~satisfied;\quad \xmark~=~not satisfied.
    \end{tablenotes}
  \end{threeparttable}
  \caption{Motivating example. The table
    reports whether~[V.1], [V.2], and~[V.2a] hold for a
    variable, and whether it is \CFInputVar{} or \recordInputVar{}.}
  \label{fig:iscfd-example}
\end{figure}

A variable that satisfies \textbf{[V.1]} and \textbf{[V.2]} is an input-dependent control-flow dependent (\CFInputVar) variable. A variable that satisfies \textbf{[V.1]} and \textbf{[V.2a]}, a strict subset of \CFInputVar, is called an input-sourced control-flow dependent (\recordInputVar) variable. 
KS-CFA records the inputs read at \recordInputVar assignments; the broader \CFInputVar values are reconstructed from these during replay.
Fig.~\ref{fig:iscfd-example} sketches an example. Offline, each direct external-input assignment of an \recordInputVar variable is marked according to whether the input is supplied by \verifier or recorded by \prover at runtime. Depending on the deployment scenario, \verifier may supply all, some, or none of the program inputs; the \verifier-versus-\prover choice of each input depends on the specific scenario. Other assignments to the variable, from constants or computation, are not \verifier's concern at runtime; \verifier derives those values from program semantics during replay. Prior work such as BDFCFA\cite{BDFCFA} requires \verifier to supply all program inputs. KS-CFA imposes no such requirement: \verifier needs only the inputs that influence control-flow transfers, whether \verifier supplies them or \prover records them at runtime.

Recording instrumentation is inserted at each direct external-input assignment whose input is sourced from \prover. At runtime, the trampoline forwards the value to the measurement engine inside the TEE, which appends it to an input log carried in the attestation report.  After \verifier receives the report,  \verifier resolves the relevant operands using program semantics or these inputs (provided by \verifier or recorded by \prover in the input log). KS-CFA avoids prover-side variable protection entirely. Prior work such as OAT \cite{OAT} (within annotated regions) and vCFI \cite{vCFI} protect all variables satisfying \textbf{[V.1]}, incurring overhead proportional to how frequently those variables are defined and used. KS-CFA's input recording, in contrast, is assignment-based: only the direct external-input assignments whose input is sourced from \prover trigger recording. It targets only \recordInputVar variables (\textbf{[V.1]} and \textbf{[V.2a]}), a strict subset of the variables targeted by OAT and vCFI.

% -----------------------------------------------------------
\subsection{Single-Path Abstract Execution} \label{sec:abstract-exec}
% -----------------------------------------------------------

Abstract execution is KS-CFA's core verification mechanism. Given a path reported by \prover, \verifier replays it symbolically, BBL by BBL, predicting the legitimate successor at every control-flow transfer and comparing it with the reported one. This section describes the offline static analysis that pre-computes the replay data (\S\ref{sec:static-analysis}), the runtime analyser that performs the replay (\S\ref{sec:runtime-analyser}), and the predictor (\S\ref{sec:predictor}).

\subsubsection{Static Analysis and VRT Construction} \label{sec:static-analysis}

Static analysis~(step~\numicon{2}) runs once per program in the offline phase. It analyses every BBL in the LLVM~IR and produces VRTs that cache the information the runtime analyser would otherwise have to re-derive from raw LLVM~IR on every attestation. For each BBL, the static analyser extracts three categories of information.

\begin{figure}
\begin{minted}[breaklines,
    autogobble,
    fontsize=\small,
    frame=lines,
    framesep=4pt]{llvm}
%a = alloca i32, align 4
store i32 10, ptr %a, align 4
%0 = load i32, ptr %a  ; -> 10
%1 = add i32 %0, 1     ; -> 11
%2 = load i32, ptr %b  ; -> load(%b)
%3 = add i32 %2, 2     ; -> add(load(%b), 2)
\end{minted}
\caption{Example of concrete evaluation.}
\label{fig:conc-eval}
\end{figure}

\emph{Abstract values of local definitions.} The analyser scans each BBL top-to-bottom, evaluating every SSA definition. Within a single BBL, it resolves expressions to concrete values when possible and records symbolic expressions when values depend on definitions from other BBLs. Consider the example in Fig.~\ref{fig:conc-eval}. Definitions \texttt{\%0} and \texttt{\%1} are resolved concretely because \texttt{\%a} is allocated and written within the same BBL. Definition \texttt{\%2} depends on \texttt{\%b}, which originates in another BBL, so the analyser records the symbolic expression \texttt{load(\%b)}; \texttt{\%3} is recorded analogously. These symbolic expressions are evaluated during abstract execution once the runtime analyser has accumulated state across multiple BBLs. Expressions that do not involve external input are resolved to concrete values, while those that depend on external input remain symbolic; expressions involving the values of \recordInputVar variables are resolved by the runtime analyser. 
Values from external input (e.g.\ \texttt{scanf}) receive a distinguished \emph{unknown} marker. For a direct external-input assignment of an \recordInputVar variable, the marker also carries the input source, so the runtime analyser can substitute the concrete value from \verifier-supplied inputs or \prover's input log during abstract execution (\S\ref{sec:runtime-analyser}). Global variables are identified with their initialisers but deferred to the runtime analyser, since their values depend on the sequence of previously executed BBLs.

\emph{Memory operations.} The analyser records every memory-modifying statement (e.g., \texttt{store}, \texttt{memcpy}, \texttt{memset}) per BBL, in execution order, so the runtime analyser can replay them. 
Within the single-BBL scope, it also emulates these operations to evaluate the abstract values that depend on them; this emulated memory state is used only for that evaluation and is not retained.
Some of these statements, however, have symbolic pointer operands that the analyser cannot resolve within a single-BBL scope. For example, an SSA computation such as \texttt{getelementptr} with a symbolic index produces a symbolic address; the subsequent \texttt{store} using that address cannot be resolved. The store is still recorded, but flagged as a \textit{deferred memory operation}: the runtime analyser evaluates the index from the runtime memory model and replays the operation. Subsequent SSA definitions or memory-modifying statements that depend on a deferred memory operation are assigned a temporary marker during static analysis. These are replaced by a concrete value or a symbolic expression after the runtime analyser replays the operation.

\emph{Control-flow metadata.} For each BBL, the analyser records: (1)~the terminator instruction (\texttt{br}, \texttt{switch}, \texttt{indirectbr}, \texttt{call}/\texttt{invoke}, \texttt{ret}) and, for multi-successor terminators, the relevant operands; (2)~any intra-program call instructions, including callee, arguments, result name, and return type; and (3)~the program entry BBL. A preliminary LLVM~IR pass splits BBLs after every intra-program call, ensuring at most one such call per BBL.

\subsubsection{Runtime Analyser} \label{sec:runtime-analyser}

Online, the runtime analyser receives the VRTs, and a control-flow path $T = \langle B_1, \dots, B_K \rangle$ expressed as LLVM~IR BBL identifiers, and two chronological input lists, $VI$ and $PI$. $VI$ holds the inputs \verifier supplied to \prover; $PI$ holds the inputs \prover captured at the instrumented assignments during execution. These cover every external input the program read on the executed path. Algorithm~\ref{alg:verifier procedure} gives the full procedure. 
The analyser performs abstract execution, which is a single-path symbolic replay of the reported trace. Unlike classical symbolic execution, it does not explore multiple paths; it follows exactly the sequence reported by \prover. The analyser maintains the runtime memory model (RMM) comprising:
\begin{itemize}
    \item A \emph{call-frame stack} (\textit{envStack}): the analyser pushes a frame on function entry and pops one on return;
    \item A \emph{per-frame environment} (\textit{env}): maps local SSA names to their current abstract values;
    \item An \emph{abstract memory} (\textit{mem}): stores the contents of objects reached via LLVM pointers. 
\end{itemize}

For each BBL, the analyser updates \textit{env} using the pre-computed abstract values from the VRT, resolving any symbolic expressions whose operands have since become concrete. It then replays the memory operations to update \textit{mem}. SSA definitions within a BBL never depend on memory writes performed \emph{in the same BBL}, because any load not resolved within the BBL 
was captured as a symbolic expression during static analysis and is evaluated through \textit{mem} during the replay phase. When the RA encounters a direct external-input assignment to an \recordInputVar variable, it reads the source attribution recorded by the static analyser, pops the next entry from \textit{VI} or \textit{PI}, and replaces the unknown marker with that value; if the indicated list is empty at that point, an anomaly is reported. Once substitution provides concrete \recordInputVar values, other LLVM variables whose values depend on them are evaluated. This makes the operands of every control-flow transfer instruction concrete, since each operand's data-dependence chain terminates at \recordInputVar variables or at values determined by program semantics.
After the update, the analyser invokes the predictor (\S\ref{sec:predictor}). Replay proceeds until the call-frame stack empties. If the trace is exhausted before the stack empties, an anomaly is reported. The same applies after replay completes if BBLs remain or either input list is non-empty.

\begin{algorithm}[!t]
\DontPrintSemicolon
\SetKwInOut{Input}{Input}\SetKwInOut{Output}{Output}
\SetKw{KwRet}{return}
\caption{Online verification procedure}
\label{alg:verifier procedure}

\Input{$T=\langle B_1,\dots,B_K\rangle$ \hfill (reported trace)\\
       $\text{VRT}$ \hfill (variable-resolution tables)\\
       $PI$  \hfill (prover-recorded inputs) \\
       $VI$ \hfill (verifier-supplied inputs)
       }
\Output{\textsc{Accepted} $\mid$ \textsc{Anomalous}}
$\textit{cur} \leftarrow \textsf{pop\_front}(T)$\;
\If{$\textit{cur} \neq$ entry BBL}{\KwRet \textsc{Anomalous}}
$\textit{envStack}\leftarrow[\;\textsf{frame}(\textit{cur})\;]$\;
$\textit{mem}\leftarrow\emptyset$\;

\While{$\textit{envStack}\neq\emptyset$}{
    $f\leftarrow\textbf{top}(\textit{envStack})$\;
    \textbf{updateRMM}$(f.\textit{env},\ \textit{mem},\ \text{VRT}[\textit{cur}],\ \textit{PI},\ \textit{VI})$\;

    \tcp{--- predict successor ---}
    $\textit{cfti}\leftarrow\textbf{getCFTI}(\textit{cur})$\;
    $k\leftarrow\textbf{classify}(\textit{cfti})$\;
    
    \uIf(\tcp*[f]{fixed target}){$k\in\{\textsf{ujmp},\textsf{dcall}\}$}{
        $p\leftarrow\textbf{fixedTarget}(\textit{cfti})$\;
    }\uElseIf{$k=\textsf{ret}$}{
        $p\leftarrow f.\textit{returnBBL}$\;
        \lIf{$T$ is empty}{\KwRet \textsc{Anomalous}}
        $r\leftarrow\textsf{pop\_front}(T)$\;
        \lIf{$p\neq r$}{\KwRet \textsc{Anomalous}}
    }
    \Else(\tcp*[f]{cjmp, ijmp, icall}){
        \lIf{$T$ is empty}{\KwRet \textsc{Anomalous}}
        $r\leftarrow\textsf{pop\_front}(T)$\;
        $p\leftarrow\textbf{concreteEval}(f.\textit{env}, \textit{mem}, \textit{cfti})$\;
        \lIf{$p\neq r$}{\KwRet \textsc{Anomalous}}
    }
    
    \tcp{--- maintain call stack ---}
    \uIf{$k\in\{\textsf{dcall},\textsf{icall}\}$}{
        $\textit{envStack}.\textsf{push}(\textsf{frame}(p))$\;
        $\textbf{top}(\textit{envStack}).\textit{returnBBL}\leftarrow\textit{cur}\texttt{.return}$\;
    }\ElseIf{$k=\textsf{ret}$}{
        $\textit{envStack}.\textsf{pop}()$\;
    }
    $\textit{cur}\leftarrow p$\;
}
\uIf{$T\neq\emptyset \lor VI\neq\emptyset \lor PI \neq\emptyset$}{\KwRet \textsc{Anomalous}}
\Else{\KwRet \textsc{Accepted}}
\end{algorithm}

\subsubsection{Prediction} \label{sec:predictor}

After updating the RMM for the current BBL, the RA retrieves the control-flow transfer instruction (\emph{cfti}) and invokes the \emph{predictor} to determine the expected successor. The predictor's behaviour depends on the transfer type:

\begin{itemize}
    \item \texttt{ujmp / dcall.} 
    The target is encoded in the IR; the predictor returns it directly. These transfers are not recorded by \prover, so no comparison is needed.
    \item \texttt{Ret.} 
    The predictor retrieves the return BBL stored when the corresponding call frame was pushed, enforcing call--return pairing. A mismatch with the reported trace signals a return-address attack.
    \item \texttt{cjmp / ijmp / icall.}
    The predictor evaluates the branch condition or target expression over the current RMM. Because \verifier has every control-flow-influencing input value (either supplied by \verifier or recorded in \prover's input log during execution), all relevant operands are concrete and evaluation yields a unique successor; a mismatch with the reported trace signals an attack.
\end{itemize}

\emph{Why the predictor’s verdict is sound.} The predictor needs the value of every \textit{cfti} operand. Some are determined by program semantics along the executed path, derivable from constants and the program's own computation. Others are \CFInputVar{} variables, whose values depend on external inputs. \verifier closes this gap with its input lists: every external input that influences a control-flow transfer, directly or indirectly, is either \verifier-supplied (carried in \textit{VI}) or \prover-recorded (carried in \textit{PI}). After substitution (\S\ref{sec:runtime-analyser}), the RA resolves the \CFInputVar{} variables using these inputs, and every \textit{cfti} operand becomes concrete.  An attacker who wishes to bend control-flow must tamper with a variable that influences a control-flow transfer (\textbf{[V.1]}). Tampering changes the variable’s in-memory value on \prover and thus alters the control-flow path that \prover reports. 
\verifier's abstract execution, however, produces its correct value by reconstructing the variable from the inputs and program semantics; \verifier's predicted edge therefore differs from the tampered reported edge, and the mismatch is detected as an anomaly. KS-CFA therefore does not require runtime protection on these variables; \verifier's abstract execution always recovers their correct values.

%-----------------------------------------------------------
\subsection{Measurement Infrastructure} \label{sec:measurement}
% -----------------------------------------------------------

The abstract-execution procedure operates on LLVM~IR BBL identifiers, but \prover captures raw runtime information. This section describes how control-flow data is captured, translated, and buffered.

\subsubsection{Instrumentation and Path Format} \label{sec:instrumentation}

\verifier instruments \prover's binary~(\numicon{6}) to capture four transfer types:  
\texttt{cjmp}, \texttt{ijmp}, \texttt{icall}, and \texttt{ret}. \texttt{ujmp} and \texttt{dcall} are not recorded because their targets are fixed in the program text and cannot be redirected without violating W$\oplus$X.
For \texttt{ijmp}, \texttt{icall}, and \texttt{ret}, the captured data is the \emph{target runtime address}; for \texttt{cjmp}, a single bit records branch-taken or branch-not-taken, since both targets are known statically.

\emph{Why runtime addresses, not BBL identifiers.} 
Reporting BBL identifiers would require instrumentation at both the entry and exit of every BBL, because \prover captures only a subset of transfer types and must still identify which BBL was entered. By contrast, capturing control-flow transfer information at the point of transfer requires instrumentation only at exits with a recorded transfer, yielding fewer capture points per BBL and less time spent in the trampoline per execution.
Additionally, runtime addresses detect \emph{mid-BBL entry attacks}: if an attacker redirects control-flow to an instruction in the interior of a BBL, the captured address will not match any legitimate position, namely a BBL start, a post-call return site, or the entry hook (the capture recording the program’s first executed BBL). This reveals the attack.
BBL identifiers, being coarser-grained, would lose this information.

\subsubsection{Address-to-BBL Translation} \label{sec:addr-bbl}

The Addr-BBL mapper translates the raw control-flow path into LLVM~IR BBL identifiers. It relies on a three-level mapping---LLVM~IR BBLs, machine BBLs, and runtime addresses---constructed by three offline analysis passes:
\begin{enumerate}
    \item \textbf{LLVM~IR analysis}~(\numicon{3}--\numicon{4}) embeds BBL names as metadata in the assembly output.
    \item \textbf{Assembly analysis}~(\numicon{7}) maps each machine BBL to its LLVM~IR BBL and records it in an ELF section.
    \item \textbf{Binary analysis}~(\numicon{9}) augments the mapping with start/end addresses, post-call addresses, and terminator types, producing a map $\textit{addr\_to\_mbb}$.
\end{enumerate}

\noindent
Algorithm~\ref{Addr-BBL_mapper algorithm} gives the translation procedure. For each element $c$ in the raw path: if $c$ is a runtime address, the mapper looks it up in $\textit{addr\_to\_mbb}$, verifies that it corresponds to a BBL start address, and emits the corresponding LLVM~IR BBL identifier. Post-call addresses receive a \texttt{.return} suffix to prevent an attacker from conflating a return landing with a normal BBL entry. 
The entry hook similarly receives a \texttt{.start} suffix to prevent an attacker from conflating the program’s initial entry with a normal entry to the entry BBL, which the program may also reach at runtime.
Any address that matches no legitimate position indicates a mid-BBL entry attack. If $c$ is a branch bit, the mapper follows unconditional control-flow edges (\texttt{ujmp}, fall-throughs, and \text{dcall} to functions in the program) from the current machine BBL until it reaches a control-flow divergence. Since branch bits represent only \texttt{cjmp}, the divergence must be a \texttt{cjmp}; reaching any other divergence (\texttt{ijmp}, \texttt{icall}, \texttt{ret}) indicates a control-flow attack. The mapper then emits the successor using the branch bit.

\begin{algorithm}[!t]
\DontPrintSemicolon
\SetKwInOut{Input}{Input}\SetKwInOut{Output}{Output}
\SetKwFunction{Lookup}{Lookup}
\SetKwFunction{LookupTgt}{TargetMBB}
\SetKwFunction{Start}{StartAddr}
\SetKwFunction{IRBBL}{IRBBL}
\SetKwFunction{IsAddr}{IsRuntimeAddr}
\SetKwFunction{Report}{ReportViolation}
\SetKwFunction{Append}{Append}
\SetKwFunction{Delegate}{DelegateMBB}
\SetKwFunction{TermType}{TermType}

\Input{$P_R$ (raw control-flow path), $\textit{addr\_to\_mbb}$}
\Output{LLVM~IR BBL sequence $T$, or \textsc{Violation}}

\BlankLine
$T \leftarrow \langle\rangle$\;
\ForEach{$c \in P_R$}{
  \uIf{\IsAddr{$c$}}{
    $\textit{mbb} \leftarrow$ \Lookup{$\textit{addr\_to\_mbb},\, c$}\;
    \lIf{$\textit{mbb}$ not found}{\Report{}}
    \uIf{$c = $ \Start{$\textit{mbb}$}}{
      \Append{$T$, \IRBBL{$\textit{mbb}$}}\;
    }
    \uElseIf{$\textit{mbb} = \textit{ENTRY\_HOOK}$ \textbf{and} $c$ is first in $P_R$}{
      \Append{$T$, \IRBBL{$\textit{mbb}$}\,
      \texttt{+\,".start"}}\;
    }
    \uElseIf{$c$ is a known post-call address in $\textit{mbb}$}{
      \Append{$T$, \IRBBL{$\textit{mbb}$}\,\texttt{+\,".return"}}\;
    }
    \Else{
      \Report{}
    }
    $\textit{cur} \leftarrow \textit{mbb}$\;
  }
  \Else(\tcp*[f]{$c$ is a branch bit}){
    $\textit{cur} \leftarrow$ \Delegate{$\textit{cur}$}\;
    \lIf{\TermType{$\textit{cur}$} $\neq$ cjmp}{\Report{}}
    $\textit{mbb} \leftarrow$ \LookupTgt{$\textit{addr\_to\_mbb},\, c,\, \textit{cur}$}\;
    \Append{$T$, \IRBBL{$\textit{mbb}$}}\;
    $\textit{cur} \leftarrow \textit{mbb}$\;
  }
}
\caption{Addr-BBL mapper}
\label{Addr-BBL_mapper algorithm}
\end{algorithm}

\subsubsection{Batch Logging} \label{sec:batch-logging}
A world switch per control-flow transfer is computationally expensive; $\approx$100--1000$\times$ slowdown has been reported
elsewhere for whole-program attestation~\cite{BLAST}. To ameliorate this, KS-CFA uses batch logging, where the trampoline accumulates control-flow data in a shared-memory buffer and triggers a world switch only when the buffer is full or the target program terminates. Because the buffer resides in non-secure memory between flushes, this enables some attack vectors. \emph{Buffer writes from program code} are blocked by software fault isolation: every store in the instrumented program is preceded by a mask-and-check sequence that rejects writes into the buffer region; the buffer boundaries are stored in read-only pages (using \texttt{mprotect}). \emph{Write-pointer corruption} is prevented by reserving a dedicated register $R_2$ for the pointer; under W$\oplus$X (\S\ref{sec:threat-model}), only existing trampoline instructions can modify $R_2$: the pointer is incremented after each log entry and reset only at step (3) of the flush routine below. KS-CFA is flexible in enabling per control-flow transfer measurements without batch logging; we explore
the performance overhead in~\S\ref{sec:eval-nobatch}.  The remaining attack is one that jumps to the trampoline's pointer-reset instruction, causing the next batch to overwrite previously logged data. KS-CFA neutralises this with a \emph{guard register}~$R_1$. The flush routine executes: (1)~set $R_1 \leftarrow 1$; (2)~world-switch to flush the buffer; (3)~reset $R_2$; (4)~check $R_1 = 1$, reporting an attack if not; (5)~set $R_1 \leftarrow 0$. No other BBL modifies $R_1$ or resets $R_2$. Jumping to step~2 and 3 is detected at step~4 because $R_1$ was never set. Jumping to step~1 commits the existing buffer contents \emph{before} resetting the pointer, so no logged data is lost.

\section{Implementation} \label{sec:implementation}

We implement KS-CFA targeting embedded devices to demonstrate its feasibility. \verifier's static analysis and compilation pipeline run on an x86-64 host using LLVM~18.1.8~\cite{LLVM}, while \prover executes on RISC-V with Keystone TEE~\cite{KeystoneTEE}. The codebase comprises approximately 25K LOC.

\subsection{Verifier Toolchain} \label{sec:impl-verifier}

\verifier's offline phase is implemented as a sequence of LLVM passes integrated into the standard compilation pipeline. A built-in \texttt{LowerSwitch} pass first converts LLVM \texttt{switch} statements to chains of \texttt{if-else} branches, so that every multi-way branch is represented as a series of two-target conditional branches in the IR. Two custom LLVM~IR passes follow: one performs LLVM~IR-level analysis, identifying BBL names and embedding them as metadata for later stages; the other instruments the \recordInputVar variables (\S\ref{sec:iscfd}) whose values are not provided by \verifier, so that their values are recorded at runtime. A machine-level pass, inserted before the \texttt{AsmPrinter}, instruments assembly code to capture control-flow transfer information at runtime. This stage requires access to hardware-level details (registers, stack layout) unavailable at the IR level. Finally, an \texttt{AsmPrinter} handler emits the LLVM-IR-to-machine-BBL mapping into custom ELF sections, which the binary analyser reads after linking. These ELF sections are not needed at runtime and can be stripped from the deployed binary. 

In our prototype, \recordInputVar variables are identified manually using two Clang/GNU-style attributes: \texttt{\_\_attribute\_\_((annotate("PInput")))} and \texttt{\_\_attribute\_\_((annotate("VInput")))}. They indicate that the variable’s value comes from a non-\verifier source or from \verifier, respectively. The recording pass instruments only variables marked \texttt{PInput}; the static analyser reads both annotations and informs the runtime analyser of each variable’s input source. Recording fires only at assignments where a \texttt{PInput}-annotated variable receives its value directly from an external-input read; assignments from constants or computation do not trigger recording. Every \recordInputVar variable must be annotated as either \texttt{PInput} or \texttt{VInput}; the choice depends on the specific attestation and target program scenario. 

KS-CFA's runtime analyser verifies execution at the LLVM~IR BBL level, while \prover executes machine-level BBLs. This imposes three requirements at the backend:
\begin{enumerate}
    \item Each LLVM~IR BBL maps to exactly one machine BBL.
    \item Any machine BBLs introduced by the backend (e.g.\ from instruction selection or register allocation) must terminate with an \texttt{ujmp} or \texttt{dcall}---transfer types whose targets are fixed and not recorded by \prover.
    \item LLVM~IR BBLs whose terminators are recorded transfer types (\texttt{cjmp}, \texttt{ijmp}, \texttt{icall}, \texttt{ret}) must be preserved: the backend must not reorder, merge, or duplicate them.
\end{enumerate}

Backend optimisations routinely violate these requirements. For instance, tail-call optimisation replaces a call--return pair with a jump, breaking call--return matching. RISC-V linker relaxation can alter instruction sequences. We thus disable backend optimisations (Table~\ref{tab:compilation-flags}). Frontend optimisations (source code to LLVM~IR) remain enabled, as they do not affect the IR-to-machine-BBL correspondence. \S\ref{sec:limitations} discusses paths toward relaxing this constraint.
\begin{table}[t]
  \centering
  \small
  \caption{Backend compiler and linker flags.}
  \label{tab:compilation-flags}
  \renewcommand{\arraystretch}{1.15}
  \begin{tabularx}{\columnwidth}{@{} l l >{\raggedright\arraybackslash}X @{}}
    \toprule
    \textbf{Flag} & \textbf{Stage} & \textbf{Purpose} \\
    \midrule
    \texttt{-O0}                 & LLC    & Disable machine-level optimisations \\
    \texttt{-disable-tail-calls} & LLC    & Preserve call--return pairing \\
    \texttt{-fast-isel=false}    & LLC    & Consistent instruction selection \\
    \texttt{-mattr=-relax}       & LLC    & Disable RISC-V relaxation in codegen \\
    \texttt{-Wl,--no-relax}      & Linker & Disable RISC-V relaxation in linker \\
    \bottomrule
  \end{tabularx}
\end{table}

\subsection{Prover Implementation} \label{sec:impl-prover}

\prover comprises a trampoline in the non-secure world and a measurement engine inside the Keystone secure enclave. The standard Keystone SDK uses event-driven communication: the enclave issues an \texttt{OCALL} to switch to the non-secure world, which invokes a dispatch function and automatically returns to the enclave when the function completes.  We modify the \texttt{OCALL} dispatcher so that it continues target-program execution after the dispatch function returns rather than switching back to the enclave. The trampoline then writes control-flow data and input records to a shared-memory buffer and triggers world switches explicitly, either when the buffer is full (batch logging) or at program termination. This modification affects only non-secure-world code and does not alter the trusted computing base (the security monitor and secure kernel remain unchanged). 

\section{Evaluation}
\label{sec:evaluation}

This section presents the evaluation of KS-CFA, discussing the approach and results for prover- and verifier-side performance using FPGA- and QEMU-based implementations.

\subsection{Experimental Setup}
\label{sec:eval-setup}

We evaluate KS-CFA using the Embench-IoT suite~\cite{patterson2025embench} using the Keystone TEE on a NiteFury~II FPGA with a single-core Rocket CPU (RV64IMAFDC, 100\,MHz, 16\,KB L1 instruction and data caches), and QEMU~7.2.1 emulating RISC-V under Ubuntu~24.04 on an Intel Ultra~9 275HX, with 21\,GB RAM. Because Embench-IoT programs use fixed inputs for deterministic scoring, we modified each program so that a single variable reads its value directly from an external input, chosen by manual inspection to suit the program’s semantics. Because the same modification is applied uniformly across the suite, whether a program’s control-flow becomes input-dependent reflects its own structure rather than program selection; each program therefore has at most one ISCFD variable (\S\ref{sec:iscfd}).
We also add two custom programs: \texttt{state-machine}, a state-machine unit test containing \texttt{ijmp} transfers (absent in the benchmark suite), and \texttt{state-machine-P}, a variant augmented with I/O operations to increase per-BBL execution time (\S\ref{sec:eval-density}).  
Selection was also bounded by manual analyst effort in annotating \recordInputVar variables for the prototype, and identifying \CFInputVar variables (\S\ref{sec:iscfd}). In the experiments, \verifier supplies all such inputs.  The program set spans control-flow density, \CFInputVar-to-CFD ratio, and program structure (cryptographic transforms, numerical computations, and state-driven dispatch).\footnote{Scale factors (LOCAL\_SCALE\_FACTOR, CPU\_MHZ) were set to 1 so that each program executes its core computation once.} Table~\ref{tab:cf-transfer-dist} summarises the control-flow distributions.

\begin{table}[t]
  \centering
  \caption{Control-flow transfer distribution across benchmarks.}
  \label{tab:cf-transfer-dist}
  \begin{threeparttable}
  \footnotesize
  \renewcommand{\arraystretch}{1.1}
  \resizebox{\linewidth}{!}{%
  \begin{tabular}{@{} l r r r r r r r @{}}
    \toprule
    Program & ujmp & cjmp & ijmp & dcall & icall & ret
            & MT* (\%) \\
    \midrule
    \texttt{mont64}  & 1\,102  &  910 &  0 &  22 & 0 &  22 & 45.3 \\
    \texttt{crc32}   & 3\,079  & 1\,027 &  0 & 1\,028 & 0 & 1\,028 & 33.4 \\
    \texttt{MAT}     & 19\,788 & 9\,725 &  0 &   805 & 0 &   805 & 33.8 \\
    \texttt{md5}     & 6\,415  & 4\,371 &  0 &   9 & 0 &   9 & 40.5 \\
    \texttt{minver}  &   293   &  222 &  0 &  13 & 0 &  13 & 43.4 \\
    \texttt{N-AES}   & 1\,902  & 1\,001 &  0 &  13 & 0 &  13 & 34.6 \\
    \texttt{N-SHA}   &   398   &   65 &  0 &   7 & 3 &  10 & 16.2 \\
    \texttt{NSICH}   &   131   &  626 &  0 &   2 & 0 &   2 & 82.5 \\
    \texttt{st}      & 2\,018  &  709 &  0 & 610 & 0 & 610 & 33.4 \\
    \texttt{SMATE}   &    30   &   53 &  0 &  13 & 0 &  13 & 60.6 \\
    \addlinespace
    \texttt{S-MACH} &    41   &   32 & 20 &   0 & 0 &   0 & 55.9 \\
    \bottomrule
  \end{tabular}%
  }
  \begin{tablenotes}\footnotesize
    \item* A multi-target (MT) transfer has $>$1 potential target BBL.
  \end{tablenotes}
  \end{threeparttable}
\end{table}

%===============================================================================
\subsection{Prover-Side Overhead}
\label{sec:eval-overhead}
\begin{figure*}[t]
  \centering
  \subfloat[Batch vs.\ no-batch slowdown on FPGA (inc.\ final TA
            processing). Numbers above `no batch' (red) are relative to `batch log' (blue).]{%
    \includegraphics[width=0.48\textwidth]{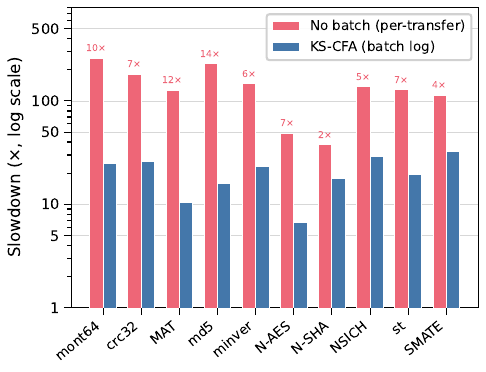}%
    \label{fig:eval-batch}%
  }\hfill
  \subfloat[Average overhead across all benchmarks, showing baseline execution, measurement overhead.]{%
    \includegraphics[width=0.48\textwidth]{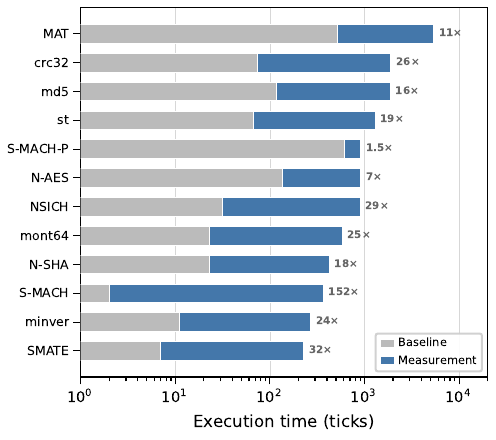}%
    \label{fig:eval-decomposition}%
  }
  \caption{Overhead results for KS-CFA on the test programs.}
  \label{fig:evaluation-overhead}
\end{figure*}

We measure runtime in two configurations: (1)~unprotected baseline, and (2)~measurement (control-flow capture + batch logging). Under the experimental setting, configuration (2) reports the measured prover-side overhead. All times exclude fixed-cost initialisation (enclave creation and trampoline setup: ${\approx}$1.07\,M~FPGA ticks, ${\approx}$1.17\,M~QEMU ticks) and finalisation (enclave destruction: ${\approx}$9\,K~FPGA ticks, ${\approx}$399\,K~QEMU ticks). Tables~\ref{tab:fpga-overhead} and~\ref{tab:qemu-overhead}, and Fig.~\ref{fig:evaluation-overhead}, report execution times and overall slowdown. Across Embench-IoT, this ranges from $2.7\times$--$18.4\times$ excluding the final TA invocation and $6.7$--$32.2\times$ including it on FPGA; the corresponding QEMU ranges are $3.8$--$13.5\times$ and $6.8$--$20.5\times$. \texttt{S-MACH} sits well outside these ranges on FPGA ($13.6\times$ excl.\ TA, $152\times$ incl.\ TA) and at the high end on QEMU ($5.8\times$ excl.\ TA, $25.5\times$ incl.\ TA); we attribute this to its very low per-BBL computation cost and analyse it separately in \S\ref{sec:eval-density}. In KS-CFA, after the target program finishes, an additional world switch is required to flush the remaining contents of the log buffer and invoke the TA to process them. As a result, there are two possible measurement endpoints: (1) program execution only, excluding the final TA processing; and (2) execution plus the final TA processing. The second naturally yields a higher overhead. 

% ---------- Table: FPGA overhead ---------------------------------------
\begin{table}[t]
  \centering
  \caption{Runtime overhead on FPGA (NiteFury~II, 100\,MHz).
           All times in ticks; Slow.\ = slowdown vs.\ baseline.}
  \label{tab:fpga-overhead}
  \begin{threeparttable}
  \footnotesize
  \renewcommand{\arraystretch}{1.05}
  \begin{tabular}{@{} l r rr rr @{}}
  \toprule
  \multirow{2}{*}[-0.8em]{Program}
    & \multirow{2}{*}[-0.8em]{\makecell{Base-\\line}}
    & \multicolumn{4}{c}{Measurement only} \\
  \cmidrule(lr){3-6}
    & & \makecell{Excl.\\TA} & \makecell{Slow.\\($\times$)}
      & \makecell{Incl.\\TA}  & \makecell{Slow.\\($\times$)} \\
  \midrule
  mont64      &    23 &    206 &  9.0 &    576 & 25.1 \\
  crc32       &    73 &  1\,117 & 15.3 &  1\,893 & 25.9 \\
  MAT         &   511 &  2\,598 &  5.1 &  5\,388 & 10.5 \\
  md5         &   118 &    833 &  7.1 &  1\,866 & 15.8 \\
  minver      &    11 &     92 &  8.1 &    268 & 23.5   \\
  N-AES       &   135 &    360 &  2.7 &    902 &  6.7 \\
  N-SHA       &    23 &     74 &  3.2 &    421 & 18.0 \\
  NSICH       &    31 &    563 & 18.4 &    895 & 29.3   \\
  st          &    66 &    731 & 11.0 &  1\,286 & 19.4 \\
  SMATE       &     7 &     58 &  8.3 &    226 & 32.2  \\
  \addlinespace
  S-MACH\tnote{*}
              &     2 &     33 & 13.6 &    365 & 152  \\
  S-MACH-P\tnote{*}
              &   617 &    682 &  1.1 &    907 &  1.5  \\
  \bottomrule
  \end{tabular}
  \begin{tablenotes}\footnotesize
    \item[*] Custom program; not part of Embench-IoT.
    \item Excl.\ TA\,=\,program execution only;
          Incl.\ TA\,=\,plus final TA processing of remaining buffer;
          Slowdowns are computed from unrounded tick counts.
          
  \end{tablenotes}
  \end{threeparttable}
\end{table}

% ---------- Table: QEMU overhead ----------------------------------------
\begin{table}[t]
  \centering
  \caption{Runtime overhead on QEMU. All times in ticks.}
  \label{tab:qemu-overhead}
  \begin{threeparttable}
  \footnotesize
  \renewcommand{\arraystretch}{1.05}
  \begin{tabular}{@{} l r rr rr @{}}
  \toprule
  \multirow{2}{*}[-0.8em]{Program}
    & \multirow{2}{*}[-0.8em]{\makecell{Base-\\line}}
    & \multicolumn{4}{c}{Measurement only} \\
  \cmidrule(lr){3-6}
    & & \makecell{Excl.\\TA} & \makecell{Slow.\\($\times$)}
      & \makecell{Incl.\\TA}  & \makecell{Slow.\\($\times$)} \\
  \midrule
  mont64   &  2\,104 & 13\,091 &  6.2 & 22\,380 & 10.6 \\
  crc32    &    910 &  6\,460 &  7.1 & 18\,610 & 20.5 \\
  MAT      &  3\,076 & 16\,179 &  5.3 & 37\,938 & 12.3 \\
  md5      &  2\,852 & 16\,748 &  5.9 & 28\,311 &  9.9 \\
  minver   &  5\,346 & 25\,713 &  4.8 & 36\,211 &  6.8 \\
  N-AES    &  5\,795 & 29\,773 &  5.1 & 48\,255 &  8.3 \\
  N-SHA    &  6\,619 & 24\,963 &  3.8 & 45\,607 &  6.9 \\
  NSICH    & 30\,632 & 412\,530 & 13.5 & 424\,174 & 13.9 \\
  st       &  2\,524 & 13\,147 &  5.2 & 26\,134 & 10.4 \\
  SMATE    &  4\,123 & 32\,463 &  7.9 & 42\,802 & 10.4 \\
  \addlinespace
  S-MACH
           &    890 &  5\,172 &  5.8 & 22\,739 & 25.5 \\
  \bottomrule
  \end{tabular}
  \begin{tablenotes}\footnotesize
    \item Notation as in Table~\ref{tab:fpga-overhead}.
          Slowdowns are computed from unrounded tick counts and rounded for readability. 
  \end{tablenotes}
  \end{threeparttable}
\end{table}

\begin{figure}[t]
  \centering
  \includegraphics[width=\columnwidth]{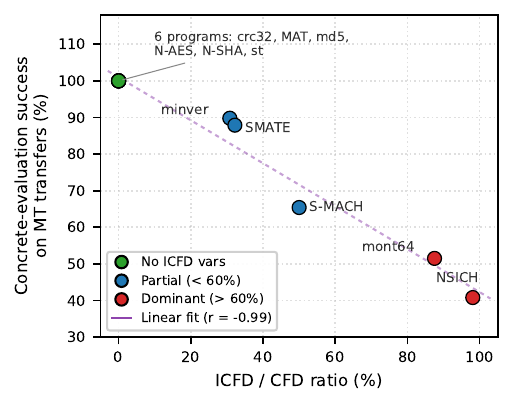}
  \caption{Concrete-evaluation success on multi-target transfers.}
  \label{fig:eval-scatter}
\end{figure}

\subsubsection{Effect of Control-Flow Density}
\label{sec:eval-density}

Runtime slowdown is dominated by control-flow density: the ratio of transfer events to useful computation per BBL. Programs whose BBLs contain only a few arithmetic instructions (e.g.\ S-MACH, $\approx$2~ticks baseline) spend most of their instrumented runtime on capture overhead, yielding extreme slowdown factors. To isolate this effect we created \texttt{S-MACH-P}, a variant of \texttt{S-MACH} that adds \texttt{printf} calls to increase per-BBL execution time and tested it on the FPGA board as primary platform. Slowdown dropped from $152\times$ to $1.5\times$ (measurement, incl.\ final~TA) 
confirming that overhead scales with control-flow density rather than with program size. The general pattern holds across all benchmarks: higher control-flow density, and a larger proportion of multi-target transfers all correlate with higher overhead.

\subsubsection{Batch Logging}
\label{sec:eval-nobatch}

Table~\ref{tab:no-batch} reports overhead when every control-flow transfer triggers an immediate world switch to the TA (i.e., batch logging disabled). Slowdown ranges from 38--260$\times$ on FPGA and 13--1\,552$\times$ in QEMU, consistent with BLAST's reproduction of C-FLAT and OAT overhead (100--1\,000$\times$)~\cite{BLAST}. On the FPGA platform, one complete measurement cycle---capture, world switch to the enclave, TA processing, and return---averages ${\sim}$7~ticks. By comparison, BLAST~\cite{BLAST} reports ARM TrustZone world-switch times across CFA schemes of $45\,\mu$s (OAT on HiKey), ${\sim}40\,\mu$s (C-FLAT on Raspberry~Pi~2), and $190\,\mu$s (BLAST on Raspberry~Pi~3). Although different hardware prevents a direct comparison, the structural difference is clear: batch logging reduces the number of switches by orders of magnitude.

% ---------- Table: no-batch overhead -------------------------------------
\begin{table}[t]
  \centering
  \caption{Overhead without batch logging (per-transfer world switch, including final TA).
           All times in ticks.}
  \label{tab:no-batch}
  \begin{threeparttable}
  \footnotesize
  \renewcommand{\arraystretch}{1.1}
  \begin{tabular}{@{} l rr rr @{}}
  \toprule
  \multirow{2}{*}{Program}
    & \multicolumn{2}{c}{FPGA}
    & \multicolumn{2}{c}{QEMU} \\
  \cmidrule(lr){2-3}\cmidrule(lr){4-5}
    & Time & Slow.\ ($\times$)
    & Time & Slow.\ ($\times$) \\
  \midrule
  mont64  &   5\,981 & 260   &   482\,443 &  229 \\
  crc32   &  13\,353 & 182   &   999\,322 & 1\,098 \\
  MAT     &  65\,138 & 127   & 4\,774\,969 & 1\,552 \\
  md5     &  27\,001 & 229   & 2\,028\,336 &  711 \\
  minver  &   1\,685 & 148   &   153\,781 &   29 \\
  N-AES   &   6\,648 &  49   &   531\,813 &   92 \\
  N-SHA   &     894 &  38   &    86\,909 &   13 \\
  NSICH   &   4\,212 & 138   &   475\,781 &   16 \\
  st      &   8\,532 & 129   &   662\,006 &  262 \\
  SMATE   &     801 & 114   &    79\,226 &   19 \\
  \bottomrule
  \end{tabular}
  \begin{tablenotes}\footnotesize
    \item Slow.\ = slowdown vs.\ unprotected baseline.
          Program abbreviations as in Table~\ref{tab:cf-transfer-dist}.
  \end{tablenotes}
  \end{threeparttable}
\end{table}

%===============================================================================
\subsection{Verification Effectiveness}
\label{sec:eval-verification}

We analyse the case in which \verifier must reconstruct the control-flow path from program semantics without program inputs. Table~\ref{tab:concrete-eval} reports the proportion of multi-target transfers \verifier resolves through concrete evaluation. 

\subsubsection{Concrete Evaluation Success}

6/11 programs (crc32, MAT, md5, N-AES, N-SHA, st) achieve 100\% concrete evaluation on multi-target transfers (\texttt{cjmp}, \texttt{ijmp}, \texttt{icall}, \texttt{ret}), independent of \verifier's input knowledge. Their control-flow is determined entirely by compile-time constants, so abstract execution yields the unique legal control-flow path without offline enumeration. Most hash-based CFA schemes (e.g., C-FLAT \cite{C-FLAT}, BDFCFA \cite{BDFCFA}) do not face measurement explosion for such programs, since their execution does not depend on external inputs, and thus only one legal control-flow path exists. The scalability advantage of KS-CFA is more pronounced for the input-dependent programs discussed in \S\ref{sec:eval-discussion}. 
Among the five other programs, evaluation success on multi-target transfers ranges from 40.8\% (NSICH) to 87.9\% (SMATE). Even without input knowledge, \verifier's abstract execution uses compile-time constants, call-site parameters, and the abstract memory model to resolve a substantial fraction of multi-target transfers. \texttt{RET} in particular always succeeds: the prediction depends on the call site (tracked by \verifier from reported transfers in RA) rather than input values.

\begin{table}[t]
  \centering
  \caption{Concrete evaluation vs.\ unresolved transfers.
           All transfers where the predictor must resolve a
           multi-target branch.}
  \label{tab:concrete-eval}
  \resizebox{\linewidth}{!}{%
  \begin{threeparttable}
  \small
  \renewcommand{\arraystretch}{1.05}
  \begin{tabular}{@{} l rr rrr rr rr @{}}
  \toprule
  Program
    & \makecell{cjmp\\conc.}
    & \makecell{cjmp\\unres.}
    & ijmp & icall & ret
    & \makecell{Total\\conc.}
    & \makecell{Total\\unres.}
    & \makecell{C/U\\(\%)}
    & \makecell{C/U\\MT (\%)} \\
  \midrule
  mont64  &   458 & 452 & --- & --- &     22 & 1\,604 & 452 &  78.0 & 51.5 \\
  crc32   & 1\,027 &   0 & --- & --- &  1\,028 & 6\,162 &   0 & 100   & 100  \\
  MAT     & 9\,725 &   0 & --- & --- &    805 & 31\,123 &   0 & 100   & 100  \\
  md5     & 4\,371 &   0 & --- & --- &      9 & 10\,804 &   0 & 100   & 100  \\
  minver  &   198 &  24 & --- & --- &     13 &    517 &  24 &  95.6 & 89.8 \\
  N-AES   & 1\,001 &   0 & --- & --- &     13 &  2\,929 &   0 & 100   & 100  \\
  N-SHA   &    65 &   0 & --- &   3 &     10 &    483 &   0 & 100   & 100  \\
  NSICH   &   254 & 372 & --- & --- &      2 &    389 & 372 &  51.1 & 40.8 \\
  st      &   709 &   0 & --- & --- &    610 &  3\,948 &   0 & 100   & 100  \\
  SMATE   &    45 &   8 & --- & --- &     13 &    101 &   8 &  92.7 & 87.9 \\
  \addlinespace
  S-MACH\tnote{*}
          &    14 &  18 &  20 & --- &      0 &     75 &  18 &  80.7 & 65.4 \\
  \bottomrule
  \end{tabular}
  \begin{tablenotes}\footnotesize
    \item[*] Custom program; not part of Embench-IoT.
    \item conc.\,=\,resolved by concrete evaluation;
          unres.\,=\,not resolved by concrete evaluation;
          C/U\,=\,concrete / (concrete + unresolved);
          MT\,=\,multi-target transfers only;
          ---\,=\,transfer type absent in program.
          Columns for ujmp and dcall are omitted (always resolved
          concretely by definition).
          ijmp, icall, and ret (where present) are all resolved by concrete evaluation; unresolved transfers are exclusively cjmp.
          Program abbreviations as in Table~\ref{tab:cf-transfer-dist}.
  \end{tablenotes}
  \end{threeparttable}
  }
\end{table}

The success rate of concrete evaluation depends on how external inputs influence control-flow transfers in the program and on the taken path. Static analysis alone cannot determine the exact rate, but the \CFInputVar-to-CFD ratio (Table \ref{tab:icfd-dist})---the share of control-flow-dependent variables that are input-dependent---serves as a useful predictor. When external inputs influence a larger share of control-flow dependent variables, the operands of control-flow transfers are more likely to depend on those inputs and therefore to remain symbolic during abstract execution. Table~\ref{tab:icfd-dist} reports this data. Six programs have zero \CFInputVar variables (external inputs do not influence their control-flow), while NSICH has 254 \CFInputVar variables out of 259 CFD variables (98\%).
These counts reflect the single external input per program (\S\ref{sec:eval-setup}): the one \recordInputVar variable is the recorded source, and the broader \CFInputVar set comprises the control-flow-relevant variables that may derive from it under static data-dependence analysis. \recordInputVar is thus the small set \verifier records, from which it reconstructs the \CFInputVar values during replay (\S\ref{sec:iscfd}).
Fig.~\ref{fig:eval-scatter} quantifies the relationship. Concrete-evaluation success on multi-target transfers follows the \CFInputVar/CFD ratio with a strong negative correlation ($r=-$0.99), and coverage declines as the \CFInputVar share rises. Even at NSICH's extreme \CFInputVar share of 98.1\%, \verifier resolves 40.8\% of multi-target transfers through abstract execution alone. The high 87.5\% headline \CFInputVar/CFD ratio of mont64 is offset by frequent loop counters that resolve concretely, so mont64 still achieves 51.5\% (Table~\ref{tab:concrete-eval}).

%------------ Table: ICFD distribution (new design)-----------
\begin{table}[t]
  \centering
  \caption{Per-program \recordInputVar, \CFInputVar, and CFD variable counts.}
  \label{tab:icfd-dist}
  \begin{threeparttable}
  \footnotesize
  \renewcommand{\arraystretch}{1.1}
  \begin{tabular}{@{} l rrr r @{}}
  \toprule
  Program
    & \recordInputVar
    & \CFInputVar
    & CFD\tnote{a}
    & \makecell{\CFInputVar/\\CFD (\%)} \\
  \midrule
  \texttt{mont64}  & 1 &  42 &  48 & 87.5  \\
  \texttt{crc32}   & 0 &   0 &   3 &  0    \\
  \texttt{MAT}     & 0 &   0 &   7 &  0    \\
  \texttt{md5}     & 0 &   0 &  18 &  0    \\
  \texttt{minver}  & 1 &   8 &  26 & 30.8  \\
  \texttt{N-AES}   & 0 &   0 &  24 &  0    \\
  \texttt{N-SHA}   & 0 &   0 &  16 &  0    \\
  \texttt{NSICH}   & 1 & 254 & 259 & 98.1  \\
  \texttt{st}      & 0 &   0 &   6 &  0    \\
  \texttt{SMATE}   & 1 &  28 &  87 & 32.2  \\
  \addlinespace
  \texttt{S-MACH}\tnote{b}
                   & 1 &   2 &   4 & 50.0  \\
  \bottomrule
  \end{tabular}
  \begin{tablenotes}\footnotesize
    \item[a] CFD\,=\,control-flow dependent variables (satisfying requirement~V.1 only; \S\ref{sec:iscfd}).
    \item[b] Custom program; not part of Embench-IoT.

    \item Program abbreviations as in Table~\ref{tab:cf-transfer-dist}.
  \end{tablenotes}
  \end{threeparttable}
\end{table}

\verifier's evaluation success depends on the input information it receives. Three benchmark programs illustrate this gradient. In md5, control-flow depends not on input content but on input \emph{length}, which is hardcoded in the program; if the length were not hardcoded, three additional variables would become \CFInputVar{} and concrete evaluation would no longer reach 100\%. Similarly, nettle-aes and nettle-sha256 would require six and two additional \CFInputVar{} variables respectively if \verifier lacks knowledge of input size. In minver, the function \texttt{mmul} has branches that depend on call-site parameters; because the caller passes hardcoded values, those variables remain non-\CFInputVar{} and concrete evaluation succeeds.  A subtler case arises in the heap-allocation function used by md5 (defined in a shared library file). A pointer derived from \texttt{heap\_ptr + size} governs a conditional branch:

\begin{minted}[fontsize=\footnotesize, autogobble]{c}
  void *next_heap_ptr = heap_ptr + size;
  if ((next_heap_ptr % sizeof(void *)) != 0) { ... }
\end{minted}

Here, \texttt{size} is a compile-time constant, and \texttt{heap\_ptr} is in a statically-defined buffer. \verifier's abstract memory model represents \texttt{heap\_ptr} symbolically with a known alignment property, so the branch outcome is determined from \texttt{size} alone. This demonstrates that the abstract memory model contributes directly to successful concrete evaluation.

These observations show that providing \verifier with as much input information as possible reduces the \CFInputVar{} variable count and increases concrete-evaluation success. Where supplying inputs is not feasible, hardcoding values in the program reduces the \CFInputVar{} variable count, thus potentially increasing concrete-evaluation success. If \verifier knows all inputs that influence control-flow transfers, \verifier can reconstruct the unique legal path  through concrete evaluation.

\subsubsection{Fallback Chain} \label{sec: fallback chain}

When concrete evaluation cannot resolve a multi-target transfer, \verifier can apply fallback mechanisms specific to the transfer type before trusting the reported edge. For \texttt{ijmp}, \verifier can use the \texttt{indirectbr} candidate list. LLVM IR's \texttt{indirectbr} instruction enumerates the potential targets of an indirect branch, and the static analyser extracts this list during the offline phase. In abstract execution, when the predictor cannot resolve the transfer due to symbolic operands, \verifier checks whether the reported target appears in the list; a target outside the list signals an attack. For \texttt{icall}, \verifier can use a conventional static CFG, derived offline, to serve as a baseline filter. When concrete evaluation on \texttt{icall} fails, \verifier checks that the reported edge is CFG-legal. \texttt{RET} is not part of the fallback chain because \verifier's call-return tracking resolves \texttt{ret} targets regardless of input knowledge.
\verifier could further reduce its reliance on the reported edge by incorporating an SMT solver (e.g., Z3) into the predictor. The solver would infer variable constraints from reported branches.  With these inferred constraints, the predictor would resolve additional \texttt{cjmp}, \texttt{ijmp}, and \texttt{icall} transfers through concrete evaluation. 
Unlike \texttt{ijmp} and \texttt{icall}, \texttt{cjmp} has no type-specific fallback check; an unresolved \texttt{cjmp} would rely on SMT constraint inference where available, and would otherwise be trusted as reported.

\subsection{Discussion}
\label{sec:eval-discussion}

KS-CFA exhibits $\approx$$7$--$32\times$ overhead on the Embench-IoT benchmarks for the measurement-only configuration including final TA processing, which is high in absolute terms but must be interpreted against two baselines. First, without batch logging, the same workloads incur $\approx$38--1{,}552$\times$ overhead (Table~\ref{tab:no-batch}), confirming that batch logging is beneficial for TEE-based CFA schemes, consistent with~\cite{BLAST}. Second, Embench-IoT programs are computation-intensive microbenchmarks with unusually high control-flow density (few instructions per BBL). 
In relation to verification strength, KS-CFA achieves 100\% concrete evaluation across the test programs when \verifier has access to all control-flow-influencing inputs. Six programs require no input recording because their control-flow depends only on compile-time constants and yields a single legal path. Although hash-based CFA schemes would also store only one hash for these single-path programs, they cannot determine this without enumerating all possible paths from the CFG and storing the resulting hashes. 
The scalability advantage is more pronounced for the remaining five programs, whose control-flow depends on external inputs.  KS-CFA avoids this trade-off entirely; concrete evaluation supported by input recording (when \verifier doesn’t supply inputs directly) handles input-dependent programs without offline path enumeration.

%===============================================================================
\section{Security Analysis}
\label{sec:security-analysis}

This section evaluates the detection mechanism against the four CFB transfer types and traditional control-flow attacks. We test KS-CFA by reproducing the in-memory state that a successful exploit would induce, isolating the detection logic (similar to~\cite{DO-RA}). Each test is performed by tampering with target variables at runtime via the \texttt{gdb} debugger in the non-secure world of the Keystone QEMU implementation.\footnote{The default Keystone QEMU image does not include \texttt{gdb} in the non-secure world; we modified the build configuration to include it.} 
The adversary capabilities are consistent with the threat model (\S\ref{sec:threat-model}), i.e., arbitrary read/write in the application's address space, knowledge of the program's source code and static CFG, and W$\oplus$X enforcement. The tests are summarised in Table~\ref{tab:attack-examples} and described below.

Attacks~\textbf{[T.1a]}--\textbf{[T.5]} tamper control-flow dependent variables whose values are determined by program semantics, validating \verifier's abstract-execution path; \textbf{[T.6]} tampers an \CFInputVar variable whose value depends on external input, and \verifier substitutes the upstream \recordInputVar inputs so that the predicate is resolved through evaluation before the branch is reached. The pair~\textbf{[T.1a]} and \textbf{[T.1b]} additionally illustrates that the Addr-BBL mapper and abstract execution provide complementary coverage of traditional control-flow attacks: the former rejects addresses that fail to align with any legitimate BBL position, while the latter rejects edges that connect valid BBLs in ways the CFG forbids.

\subsection{Traditional Control-flow Attacks}  Conventional ROP~\cite{ROP_Original}, JOP~\cite{bletsch2011jump}, and COP~\cite{COP} attacks redirect execution to addresses that are not valid BBL entry points or that violate the program's edge structure. The Addr-BBL mapper (Algorithm~\ref{Addr-BBL_mapper algorithm}) rejects any reported address that does not match a legitimate position (a BBL start address, a recorded post-call return site, or the entry hook), and the edge-by-edge replay rejects any transition inconsistent with the program's structure. By experiment, we tamper with a return address in \texttt{mont64} under two configurations. In~\textbf{[T.1a]}, the corrupted address lands mid-BBL, matching neither a BBL start, a recorded post-call return site, nor the entry hook. Thus, the Addr-BBL mapper rejects it outright. In~\textbf{[T.1b]}, the corrupted address targets a legitimate BBL entry that is nonetheless invalid for this call site: a CFG-illegal edge whose destination coincides with a valid BBL boundary. The mapper accepts the address, but the runtime analyser records the expected return BBL and detects the mismatch during abstract execution.

\subsection{CFB-CJMP}  In this attack (\textbf{[T.2]}), an adversary corrupts a branch variable or loop counter to alter a conditional branch. 
Let $v$ be the corrupted variable and $B_i$ the BBL containing the conditional branch. If $v$ is \emph{not} \CFInputVar---i.e., its value is deterministic from the program semantics---the predictor's concrete evaluation over the RMM yields a unique expected target~$B_j$. The attacker's corruption causes \prover to report a different successor~$B_k \neq B_j$; the consistency check is triggered and the trace is flagged \textsc{anomalous}. If $v$ \emph{is} \CFInputVar, concrete evaluation requires external input. However, \verifier knows every external input that influences control flow directly or indirectly, either from \verifier's side or from \prover's side. Therefore, \verifier is able to replace the \recordInputVar variable value when assigned by the external inputs, which makes the related prediction successful and yields the unique target BBL. The attacker bends \prover's control flow, but \verifier knows the proper values of control-flow-dependent variables and computes the right target $B_j$. \prover therefore reports a different successor $B_k \neq B_j$, and the trace is flagged as \textsc{anomalous}. By experiment, we target \texttt{mont64}'s main loop. We flip the loop bound \texttt{rpt} from~1 to~0, causing the core computation to be skipped. Because \texttt{rpt} is resolvable by concrete evaluation, the predictor computes the legitimate successor and rejects the divergence.

\subsection{CFB-IJMP}
Here, the attacker targets indirect jumps. They may tamper with an index or pointer used by an indirect branch (\texttt{indirectbr}), redirecting execution to a sibling BBL within the same jump table. The analysis is structurally identical to CFB-CJMP: if the index is \verifier-known, concrete evaluation predicts the correct target and a mismatch is detected. If it is \CFInputVar, \verifier uses external inputs to make concrete evaluation succeed and yield the proper successor. In \textbf{[T.3]}, we redirect an indirect jump to a different but CFG-legal jump-table target in the \texttt{S-MACH} program, skipping a state's operations. The predictor resolves the dispatch index concretely and flags the divergence.

\subsection{CFB-ICALL}

\textbf{[T.4]} investigates CFB attacks against indirect calls. The attacker overwrites a function pointer or virtual-table entry so that an indirect call dispatches to an unintended yet CFG-legal callee. If the pointer's value is \verifier-known, the predictor resolves the call target concretely and detects the mismatch. If the pointer is \CFInputVar, \verifier resolves the call target concretely through external input substitution, and the mismatch is detected.  \textsc{N-SHA}'s existing icall sites are single-target, so to exercise multi-target indirect dispatch, we extend \textsc{N-SHA} with a second initialisation function that shares the SHA-256 context layout but loads a different initial state. The two functions are dispatched through a function-pointer table, yielding a two-target icall site that is CFG-legal under fully precise static CFI; this configuration mirrors common patterns in cryptographic libraries that select between algorithm variants at initialisation. We overwrite the function pointer at the call site; the predictor computes the legitimate target and detects the mismatch. We note that this is a methodological compromise driven by the absence of multi-target \texttt{icall} sites in the Embench-IoT programs (\S\ref{sec:eval-setup}), but the detection mechanism itself is independent.

\subsection{CFB-RET}

The attacker targets return addresses; for example, by overwriting the saved return address on the stack so that the callee returns to a different but CFG-legal caller (\textbf{[T.5]}). KS-CFA detects this without monitoring the return address directly. On \texttt{icall}, the trampoline records the dynamically resolved target; \texttt{dcall} targets are statically known to \verifier and need not be recorded. In both cases, at verification time the runtime analyser pushes the expected return BBL (the instruction following the call site, encoded as \texttt{callBBL.return}) onto a shadow call stack on entry to the callee. When a \texttt{ret} is encountered, the predictor pops the shadow stack and compares the expected return BBL against the reported successor. A corrupted return address causes the reported successor to diverge from the shadow-stack entry, and the trace is flagged \textsc{anomalous}. This enforces strict call-return correspondence without requiring a hardware shadow stack. By experiment, we redirect the return address of \texttt{mulul64}---called from multiple sites in \texttt{mont64}---to a different but CFG-legal caller's return site. Each call frame pushes the expected return BBL onto the shadow call stack, and the divergence between expected and reported successor surfaces in abstract execution.

 \subsection{\CFInputVar{} Variable Tampering}  In this experiment (\textbf{[T.6]}), we tamper with an \CFInputVar predicate in \texttt{minver\_fabs} (program \texttt{minver}) to flip the outcome of a conditional branch. Although concrete evaluation of the branch requires external input, \verifier substitutes the upstream \recordInputVar inputs during abstract execution, and the predicate is resolved through subsequent evaluation before the branch is reached. The predictor computes the unique legitimate successor and rejects the divergence.

\begin{table}[t]
  \centering
  \caption{Attack summary and the associated detection mechanisms.}
  \label{tab:attack-examples}
  \resizebox{\linewidth}{!}{%
  \begin{threeparttable}
  \footnotesize
  \renewcommand{\arraystretch}{1.1}
  \begin{tabular}{@{} r c c l l @{}}
  \toprule
  ID
    & \makecell{Attack\\Class}
    & Program
    & Corruption target
    & Detection mechanism \\
  \midrule
  \textbf{[T.1a]} & Trad.       & \texttt{mont64}            & Return address      & Addr-BBL mapper        \\
  \textbf{[T.1b]} & Trad.       & \texttt{mont64}            & Return address      & Abstract execution     \\
  \textbf{[T.2]}  & A4          & \texttt{mont64}            & Loop bound (\texttt{rpt}) & Abstract execution \\
  \textbf{[T.3]}  & A3          & \texttt{S-MACH}            & Dispatch index      & Abstract execution     \\
  \textbf{[T.4]}  & A1          & \texttt{N-SHA}    & Function pointer    & Abstract execution     \\
  \textbf{[T.5]}  & A2          & \texttt{mont64}            & Return address      & Abstract execution     \\
  \textbf{[T.6]}  & A4          & \texttt{minver}            & Predicate (\texttt{n}) & Abstract execution \\
  \bottomrule
  \end{tabular}
  \begin{tablenotes}\footnotesize
    \item Trad.\,=\,traditional CFG-violating control-flow attacks
          (ROP/JOP/COP).
    \item A1--A4 denote CFB attack types from
          Table~\ref{tab:general comparison}.  Programs as in Table~\ref{tab:cf-transfer-dist}.
          
  \end{tablenotes}
  \end{threeparttable}
  }
\end{table}

\section{Limitations and Future Work} \label{sec:limitations}
In this section, we discuss the limitations of this work and outline potential mitigations and future directions.

\subsection{Prototype Limitations}

Our prototype identifies \recordInputVar variables through manual annotations. A missed annotation leaves an external input unrecorded, creating a potential blind spot.
Over-annotated variables, in contrast, incur a small penalty. Such variables are not consulted by the predictor at a control-flow transfer; they carry no security consequence, and cost only extra recorded inputs, or additional \verifier-side storage. 
Automatically labelling relevant variables would be a fruitful direction of future research. Example approaches may involve taint analysis from external-input sources (e.g., \texttt{scanf}, socket reads, file I/O), and a data-dependence analysis that identifies variables whose values directly or transitively influence control-flow transfers, which we defer to future research.

Additionally, the static analyser marks variables that receive return values from system library calls as \recordInputVar, even when their outputs are deterministic given known parameters. This inflates the \recordInputVar set and increases \prover overhead. Extending the analyser to model common library functions would reduce this. The static analysis and abstract execution phases also currently process \emph{all} LLVM variables, including those with no influence on control-flow transfers. A pre-processing pass that eliminates such variables from the VRTs would reduce offline analysis time and online verification cost. The prototype also targets statically-linked binaries with ASLR disabled, both common simplifications in CFA implementations (e.g., OAT \cite{OAT}). 

Finally, reporting runtime addresses, branch bits, and the input log produces longer attestation logs than hash-based schemes that emit one or more compact digests per execution. For long-running programs, this increases communication and storage costs. KS-CFA already elides deterministic edges (\texttt{ujmp} and \texttt{dcall}, which are not recorded); existing orthogonal techniques, e.g., periodic flushing~\cite{ACFA} and compression~\cite{SCARR}, can mitigate this. An evaluation of trace-compression strategies is also deferred to future work.

\subsection{Compilation Constraints} \label{sec:lim-compilation}
The prototype's runtime analyser verifies execution at LLVM IR BBL granularity, but \prover executes machine BBLs; backend compiler transformations can break the correspondence between the two. To maintain a faithful mapping, the prototype enforces three compilation constraints (\S\ref{sec:implementation}), which together necessitate disabling backend optimisations and impose a performance penalty. 
A promising direction for relaxing this constraint is a binary lifter (e.g., Rellume~\cite{Rellume}), which could reconstruct LLVM IR from the final binary, producing an IR whose BBL layout more directly reflects the machine code. RISC-V's regular instruction encoding is well-suited to lifting; unlike x86, it lacks variable-length instructions and complex multi-operation primitives (e.g., repeated string instructions) that complicate lifters. An evaluation of this approach is left to future work.

\subsection{Alternative Deployments} \label{sec:lim-composability}

KS-CFA's verification approach---selective input handling and single-path symbolic replay---is not inherently tied to Keystone.
In principle, any CFA scheme whose \prover reports the targets of control-flow transfers (runtime addresses, branch bits, or equivalent) could integrate KS-CFA's replay engine as an additional verification layer, though adaptations would be needed to handle different path formats (e.g., machine-level BBL identifiers, or paths that include unconditional transfers). 
As discussed in \S\ref{sec:eval-verification}, even when the prover cannot provide program inputs (e.g., for privacy), KS-CFA can still detect CFB attacks on its own. In this case, when KS-CFA is layered on conventional CFA schemes, not all fallback mechanisms described in \S\ref{sec: fallback chain} are applicable, but static CFG integration is generally available to detect traditional control-flow attacks. For binary-based conventional schemes, KS-CFA could potentially use the binary-lifter approach discussed in \S\ref{sec:lim-compilation} to obtain the required LLVM IR. Therefore, integrating KS-CFA's verification engine with these schemes extends detection coverage to CFB attacks, a more sophisticated class of control-flow attack.

\section{Conclusion} \label{sec:conclusion}
We presented KS-CFA, a CFA scheme that addresses CFB attacks through the \recordInputVar variables concept
and single-path symbolic replay. It provides detection across four exploitable transfer types without requiring a static CFG, dedicated hardware, or verifier-side path enumeration.
We implemented KS-CFA on RISC-V using Keystone and evaluated it using Embench-IoT on both QEMU and NiteFury II FPGA platform. The concrete evaluation success can reach 100\% when \verifier has access to all control-flow-influencing inputs, meaning \verifier reconstructs the unique legal path. Prover-side overhead on the Embench-IoT benchmarks ranges from 6.7× to 32.2× on the FPGA and 6.8× to 20.5× on QEMU.
Automating \recordInputVar identification and adopting binary lifting are the principal directions for future work. More broadly, we demonstrate that selective input handling combined with single-path verification offers a practical design point for CFB resilience in embedded attestation.

% Set reference style and display
\bibliographystyle{IEEEtran}
\bibliography{refs}
\vspace{-0.5cm}
\begin{IEEEbiography}[{\includegraphics
[width=1in,height=1.25in,clip,
keepaspectratio]{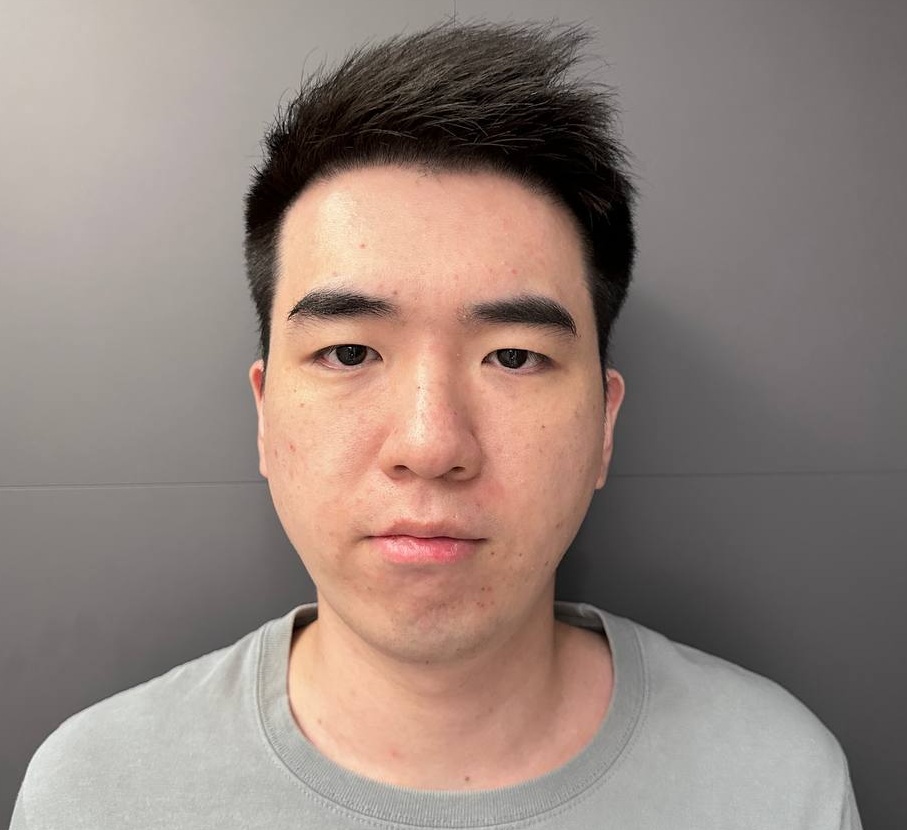}}]
{Zhanyu Sha} (B.Sc., M.Sc.) is currently a Ph.D.\ student in Information Security at Royal Holloway, University of London. He received his M.Sc.\ in Cyber Security from King's College London. He also earned two B.Sc.\ degrees: one in Information and Computer Science from Xi'an Jiaotong-Liverpool University, China, and another in Computer Science from the University of Liverpool, U.K. His research interests include software security, trusted execution environments, and embedded systems.
\end{IEEEbiography}
\vspace{-0.5cm}

\begin{IEEEbiography}[{\includegraphics
[width=1in,height=1.25in,clip,
keepaspectratio]{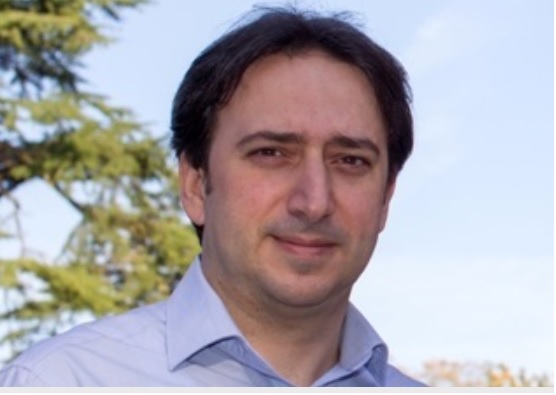}}]
{Konstantinos Markantonakis} (B.Sc., M.Sc., MBA, Ph.D.) is a Professor of Information Security in Royal Holloway University of London, and the Director of the Information Security Group Smart Card and IoT Security Centre (SCC). He obtained his B.Sc.\ (Lancaster University), M.Sc., Ph.D.\ (London) and his MBA in International Management from Royal Holloway, University of London. His research interests include smart card security and applications, secure cryptographic protocol design, embedded systems security, autonomous systems and trusted execution environments.
\end{IEEEbiography}
\vspace{-0.5cm}

\begin{IEEEbiography}[{\includegraphics
[width=1in,height=1.25in,clip,
keepaspectratio]{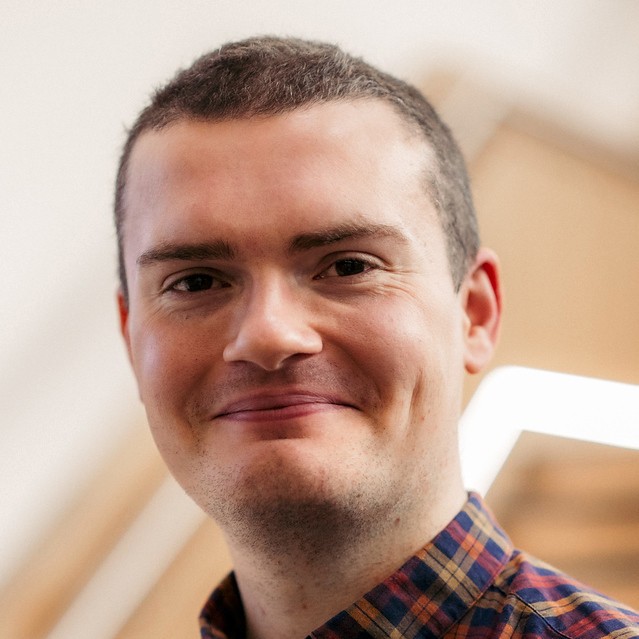}}]
{Carlton Shepherd} (B.Sc., Ph.D.) received his Ph.D.\ Information Security from Royal Holloway, University of London, and his B.Sc.\ in Computer Science from Newcastle University. He is currently Assistant Professor of Cyber Security at Durham University. Previously, he was a Lecturer in Computer Science at Newcastle University, and a Senior Research Fellow at the Information Security Group at Royal Holloway, University of London. His research interests centre around the security of trusted execution environments (TEEs) and their applications, secure CPU design, embedded systems, applied cryptography, and hardware security.
\end{IEEEbiography}
\vspace{-0.5cm}

\begin{IEEEbiography}[{\includegraphics
[width=1in,height=1.25in,clip,
keepaspectratio]{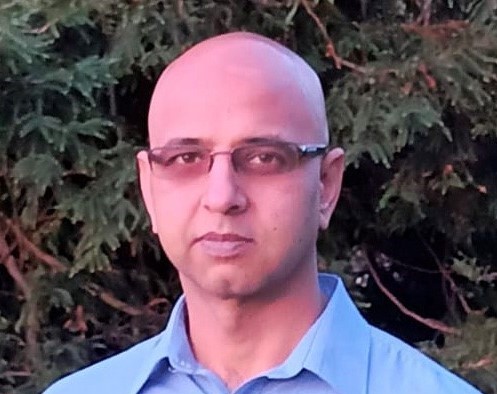}}]
{Amir Rafi} (B.Sc., M.Sc.) obtained his M.Sc.\ in Information Security from Royal Holloway, University of London, his B.Sc.\ in Computing from Queen Mary, University of London, and he is currently completing his Ph.D.\ in Information Security at Royal Holloway, University of London. He is a member of the Information Security Group Smart Card and IoT Security Centre (SCC) and was previously Research Assistant at the Information Security Group at Royal Holloway, University of London. His research interests include digital rights management, trusted execution environments and embedded systems security.
\end{IEEEbiography}

\end{document}